\documentclass[letterpaper]{article}
\usepackage[margin=1in]{geometry}
\usepackage{hyperref}
\hypersetup{
	colorlinks   = true,    
	citecolor    = red      
}
\usepackage[onehalfspacing]{setspace}

\usepackage{amsmath,mathtools}
\usepackage{subcaption}
\usepackage{booktabs}
\usepackage[braket, qm]{qcircuit}

\usepackage{pgfmath}
\usepackage{amsmath}
\usepackage{xparse}

\usepackage{alltt}
\usepackage{newverbs}
\usepackage{comment}

\usepackage{setspace}
\usepackage{listings}
\usepackage{xcolor}
\usepackage{bm}

\usepackage{braket}

\usepackage{graphicx}
\usepackage{subcaption}
\begingroup
  \usepackage{physics}
  \global\let\wantedmacro\wantedmacro
  \global\expandafter\let\csname pdv \expandafter\endcsname
\endgroup
\usepackage{authblk}
\usepackage{xcolor}
\usepackage[linesnumbered,ruled,vlined]{algorithm2e}
\usepackage{qcircuit}
\usepackage{booktabs}
\usepackage{svg}
\usepackage{cite}
\usepackage[normalem]{ulem}

\usepackage{float}

\SetCommentSty{mycommfont}

\SetKwInput{KwInput}{Input}                
\SetKwInput{KwOutput}{Output}              
\SetKwProg{Fn}{Function}{}{}

\newcommand{\pbr}[2]{\lcb{#1, \, #2}}

\newcommand{\dbr}[2]{{\lcb{#1, \, #2}}_{DB}}

\newcommand{\cbr}[2]{{\lsb{#1, \, #2}}}

\newcommand{\lob}[1]{\left( #1 \right)} 
\newcommand{\lcb}[1]{\left\{ #1 \right\}} 
\newcommand{\lsb}[1]{\left[ #1 \right]} 

\newcommand{\twobyone}[2]{\begin{bmatrix} #1 \\ #2 \end{bmatrix}}

\newcommand{\twobytwo}[4]{\begin{bmatrix} #1 & #2 \\ #3 & #4 \end{bmatrix}}

\newcommand{\pdir}[2]{\frac{\partial #1}{\partial #2}} 
\newcommand{\pddir}[2]{\frac{{\partial^2 #1}}{{\partial #2^2 }}}

\newcommand{\iseqp}[1]{1,\, 2,\, \ldots\,,\, #1}

\newcommand{\mseqgset}[3]{%
   \pgfmathparse{int(#1 + #3)}%
  \text{ $ \{ #1$, $\pgfmathresult$, $\ldots \,$, $#2 \}$} %
}

\newcommand{\mseqset}[2]{%
    \pgfmathparse{}%
   \mseqgset{#1}{#2}{1}
}

\newcommand{\mseqgenum}[3]{%
   \pgfmathparse{int(#1 + #3)}%
  \text{ $  #1$, $\pgfmathresult$, $\ldots \,$, $#2 $} %
}

\newcommand{\mseqenum}[2]{%
    \pgfmathparse{}%
   \mseqgenum{#1}{#2}{1}
}

\newcommand{\meqref}[1]{\text{Eq}.~\eqref{#1}}
\newcommand{\mref}[1]{Sec.\,\,\!$\ref{#1} $}
\newcommand{\mfig}[1]{Fig.\,\,\!$\ref{#1} $}

\usepackage[english]{babel}

\usepackage{pdflscape}
\usepackage{fancyvrb}
\usepackage{calc}
\usepackage{rotating}
\usepackage{pgf,tikz}
\usepackage{mathrsfs}
\usetikzlibrary{arrows}
\usepackage{mathtools}

\usepackage{listings}
\usepackage[]{qcircuit}
\usepackage{braket}
\usepackage{mathtools}
\usepackage{varwidth}
\usepackage{caption}
\usepackage{subcaption}
\usepackage{graphicx}
\usepackage[export]{adjustbox}

\makeatletter
\def\paragraph{\@startsection{paragraph}{4}%
	\z@\z@{-\fontdimen2\font}%
	{\normalfont\bfseries}}
\makeatother

\usepackage{graphicx}
\usepackage{pgffor}
\usepackage{tikz,tikz-cd}
\usetikzlibrary{backgrounds,fit,decorations.pathreplacing}  
\usepackage{booktabs}
\usepackage{enumerate}

\usepackage{amssymb, amsmath, amsthm}

\usepackage{graphicx}              
\usepackage{amsmath}               
\usepackage{amsfonts}              
\usepackage{amsthm}                
\usepackage{xkeyval}               
\usepackage{amssymb}
\usepackage{enumerate}             
\usepackage{color}
\usepackage{breqn}                 
\usepackage{bm}                    

\allowdisplaybreaks[1]
\usepackage{nicefrac}
\usepackage{booktabs}

\usepackage{kpfonts}
\usepackage{stackengine}
\usepackage{calc}
\newlength\shlength
\newcommand\xshlongvec[2][0]{\setlength\shlength{#1pt}%
	\stackengine{-5.6pt}{$#2$}{\smash{$\kern\shlength%
			\stackengine{7.55pt}{$\mathchar"017E$}%
			{\rule{\widthof{$#2$}}{.57pt}\kern.4pt}{O}{r}{F}{F}{L}\kern-\shlength$}}%
	{O}{c}{F}{T}{S}}

\makeatletter
\let\save@mathaccent\mathaccent
\newcommand*\if@single[3]{%
  \setbox0\hbox{${\mathaccent"0362{#1}}^H$}%
  \setbox2\hbox{${\mathaccent"0362{\kern0pt#1}}^H$}%
  \ifdim\ht0=\ht2 #3\else #2\fi
  }
\newcommand*\rel@kern[1]{\kern#1\dimexpr\macc@kerna}
\newcommand*\widebar[1]{\@ifnextchar^{{\wide@bar{#1}{0}}}{\wide@bar{#1}{1}}}
\newcommand*\wide@bar[2]{\if@single{#1}{\wide@bar@{#1}{#2}{1}}{\wide@bar@{#1}{#2}{2}}}
\newcommand*\wide@bar@[3]{%
  \begingroup
  \def\mathaccent##1##2{%
    \let\mathaccent\save@mathaccent
    \if#32 \let\macc@nucleus\first@char \fi
    \setbox\z@\hbox{$\macc@style{\macc@nucleus}_{}$}%
    \setbox\tw@\hbox{$\macc@style{\macc@nucleus}{}_{}$}%
    \dimen@\wd\tw@
    \advance\dimen@-\wd\z@
    \divide\dimen@ 3
    \@tempdima\wd\tw@
    \advance\@tempdima-\scriptspace
    \divide\@tempdima 10
    \advance\dimen@-\@tempdima
    \ifdim\dimen@>\z@ \dimen@0pt\fi
    \rel@kern{0.6}\kern-\dimen@
    \if#31
      \overline{\rel@kern{-0.6}\kern\dimen@\macc@nucleus\rel@kern{0.4}\kern\dimen@}%
      \advance\dimen@0.4\dimexpr\macc@kerna
      \let\final@kern#2%
      \ifdim\dimen@<\z@ \let\final@kern1\fi
      \if\final@kern1 \kern-\dimen@\fi
    \else
      \overline{\rel@kern{-0.6}\kern\dimen@#1}%
    \fi
  }%
  \macc@depth\@ne
  \let\math@bgroup\@empty \let\math@egroup\macc@set@skewchar
  \mathsurround\z@ \frozen@everymath{\mathgroup\macc@group\relax}%
  \macc@set@skewchar\relax
  \let\mathaccentV\macc@nested@a
  \if#31
    \macc@nested@a\relax111{#1}%
  \else
    \def\gobble@till@marker##1\endmarker{}%
    \futurelet\first@char\gobble@till@marker#1\endmarker
    \ifcat\noexpand\first@char A\else
      \def\first@char{}%
    \fi
    \macc@nested@a\relax111{\first@char}%
  \fi
  \endgroup
}
\makeatother

\newcommand{\Schrodinger}{Schr\"{o}dinger}

\newcommand{\RN}[1]{%
	\textup{\uppercase\expandafter{\romannumeral#1}}%
}

\allowdisplaybreaks

\newtheorem{thm}{Theorem}[subsection]

\newtheorem{defn}[thm]{Definition} 
\newtheorem{example}[thm]{Example} 
\newtheorem{remark}[thm]{Remark}

\newcommand{\RR}{\mathbb{R}}      

\newcommand{\mat}[4]{\left[\begin{smallmatrix*}[r]
		#1 & #2 \\
		#3 & #4 \\
	\end{smallmatrix*}\right]}

 \newcommand{\bmat}[4]{\begin{bmatrix}
    \begin{array}{rr}
       #1 & #2 \\
		#3 & #4 \\
    \end{array}
\end{bmatrix}}

\newcommand{\cosbasisoned}[1] {\frac{1}{\sqrt{L_{#1}}} \cos\left(\frac{ (k-\frac{1}{2}) \pi x}{L_{#1}}\right)}
\newcommand{\sinbasisoned} [1]{ \frac{1}{\sqrt{L_{#1}}} \sin\left(\frac{k \pi x}{L_{#1}}\right)}

\newcommand{\cosbasis}[1] {\frac{1}{\sqrt{L_{#1}}} \cos\left(\frac{ (k_{#1} -\frac{1}{2}) \pi x_{#1}}{L_{#1}}\right)}
\newcommand{\sinbasis} [1]{ \frac{1}{\sqrt{L_{#1}}} \sin\left(\frac{k_{#1} \pi x_{#1}}{L_{#1}}\right)}

\def\Pp{{\cal P}}

\def\Ss{{\cal S}}

\def\Xx{{\cal X}}

\def\RR{{\mathbb R}}

\def\<{\langle}
\def\>{\rangle}

\numberwithin{equation}{section}

\makeatletter
\def\mwidetilde#1{\mathop{\vbox{\m@th\ialign{##\crcr\noalign{\kern3\p@}%
      \sortoftildefill\crcr\noalign{\kern3\p@\nointerlineskip}%
      $\hfil\displaystyle{#1}\hfil$\crcr}}}\limits}

\def\sortoftildefill{$\m@th \setbox\z@\hbox{$\braceld$}%
  \braceld\leaders\vrule \@height\ht\z@ \@depth\z@\hfill\braceru$}
\makeatother

\begin{document}

 \title{A quantum approach for optimal control}

	\author[1]{Hirmay Sandesara}
	\author[2]{Alok Shukla \thanks{Corresponding author.}}
	\author[3]{Prakash Vedula}
	\affil[1,2]{School of Arts and Sciences, Ahmedabad University, India}
	\affil[1]{hirmaysandesara@gmail.com}
	\affil[2]{alok.shukla@ahduni.edu.in}
	\affil[3]{School of Aerospace and Mechanical Engineering, University of Oklahoma, USA}
	\affil[3]{pvedula@ou.edu}

	\date{}
	\maketitle

\begin{abstract}
In this work, we propose a novel variational quantum approach for solving a class of nonlinear optimal control problems. Our approach integrates Dirac's  canonical quantization of dynamical systems with the solution of the ground state of the resulting non-Hermitian Hamiltonian via a variational quantum eigensolver (VQE). We introduce a new perspective on the Dirac bracket formulation for generalized Hamiltonian dynamics in the presence of constraints, providing a clear motivation and illustrative examples. Additionally, we explore the structural properties of Dirac brackets within the context of multidimensional constrained optimization problems.

Our approach for solving a class of nonlinear optimal control problems employs a VQE-based approach to determine the eigenstate and corresponding eigenvalue associated with the ground state energy of a non-Hermitian Hamiltonian. Assuming access to an ideal VQE, our formulation demonstrates excellent results, as evidenced by selected computational examples. Furthermore, our method performs well when combined with a VQE-based approach for non-Hermitian Hamiltonian systems.
Our VQE-based formulation effectively addresses challenges associated with a wide range of optimal control problems, particularly in high-dimensional scenarios. Compared to standard classical approaches, our quantum-based method shows significant promise and offers a compelling alternative for tackling complex, high-dimensional optimization challenges.
\end{abstract}

\section{Introduction}\label{sec:intro}
In recent years, numerous quantum and hybrid classical-quantum algorithms have been introduced \cite{nielsen2002quantum, deutsch1992rapid, bernstein1993quantum, shukla2023generalization, simon1997power, grover1996fast, shukla2024grover, wittek2014quantum, childs2020quantum, shukla2023hybrid, lloyd2020quantum, shukla2019trajectory, yan2016survey, Shukla2022, rohida2024hybrid}, spanning diverse fields and often demonstrating considerable computational advantages over classical methods.
However, the small number of
qubits, decoherence, state preparation, and measurement challenges limit the realization of quantum advantage for practical applications. As research is being carried out to tackle these challenges \cite{shukla2024efficient,  campbell2024series, van2023quantum}, 
Variational Quantum Algorithms (VQAs) \cite{cerezo2021variational} have emerged as the leading approach to harness quantum advantage on  
Noisy Intermediate-Scale Quantum (NISQ) devices. In this work, we propose a quantum approach based on Variational Quantum Algorithms for non-Hermitian systems for solving nonlinear optimal control problems.

Nonlinear optimal control problems are important in many disciplines such as engineering, natural sciences, finance and economics \cite{kirk_opt_control, lewis2012optimal}. These problems often involve determining the control inputs over a specific time horizon to optimize a performance criterion, while the system dynamics are governed by nonlinear differential equations. The importance of these problems arises from their practical applicability in real-world situations where systems often exhibit nonlinear characteristics \cite{lewis2012optimal}.

One approach for solving nonlinear optimal control problems involves formulating them as Hamiltonian systems. In this framework, the solution of the optimal control problem is equivalent to integrating a Hamilton–Jacobi equation backward in time from a specified terminal cost \cite{pontryagin1987mathematical}. However, the presence of nonlinearity poses significant challenges in obtaining analytical solutions to these equations. Various proposed approximate solutions of the Hamilton-Jacobi equation often involve time-intensive computations and provide no guarantee that resulting control variables exhibit improved performance \cite{Sakamoto}. Furthermore, traditional methods often struggle to handle the complexities introduced by nonlinear dynamics, leading to computational inefficiencies and practical implementation challenges \cite{kirk_opt_control}.

To address some of these challenges, we focus on extending the existing framework for solving nonlinear optimal control problems by framing them as quantum mechanical eigenvalue problems. This approach uses the concept of canonical quantization from quantum mechanics to efficiently address nonlinear optimal control problems.
We present a novel perspective on the Dirac bracket formulation for generalized Hamiltonian dynamics with constraints, offering clear motivation and illustrative examples. Additionally, we explore the structural properties of Dirac brackets in the context of multidimensional constrained optimization problems. 
A quantum mechanical approach for optimal control problems has also been described in other works, including \cite{itami2005nonlinear}. However, our formulation is different and more closely follows Dirac's original quantization program (as
our proposed approach carefully accounts for first-class, second-class, primary, and secondary constraints) leading to a more elegant and simpler construction of the control framework. 
For instance, the construction of Dirac brackets for a class of one-dimensional problems involves a $2 \times 2$ matrix (ref.~\meqref{eq:M} and \meqref{eq:MInv}), whereas in \cite{itami2005nonlinear}, an analogous construction involves a $4 \times 4$ matrix. Further, we provided a new definition of the Dirac bracket in the context of the class of optimal control problems using the structural properties of the matrices involved (ref.~\meqref{eq:defNewDirac}). This formulation provides a convenient approach for computation of Dirac brackets. 

Once the optimal control problem is formulated in terms of a \Schrodinger \ wave equation, spectral methods (with different basis functions) are employed to obtain a numerical solution of the wave equation. Subsequently, we get back the classical solution of the optimal control problem in the limit $h_w \to 0$, where $h_w$ is related to Planck's constant in our quantum mechanical formulation of the optimal control problem. In this work, we also explore the dependence of the quality of obtained solutions on factors such as the eigenvalues of the Hamiltonian matrix, the number of basis functions employed, and the chosen value of $h_w$.

Further, we extend our proposed framework to a multidimensional setting, accommodating more complex dynamical systems. Another important novelty of our work is integrating a modified Variational Quantum Algorithm (VQA) tailored to handle non-Hermitian Hamiltonian matrices arising from nonlinear systems with our quantum mechanical framework for optimal control problems. This modified VQA draws inspiration from recent advancements in quantum computing and optimization techniques, particularly from the approach presented in \cite{Xie2024}. Previous works \cite{itami2005nonlinear, contreras2019quantum} have been limited to obtaining numerical solutions for optimal control problems (despite their formulation as quantum mechanical problems) using classical computers. In contrast, we propose solving these optimal control problems by leveraging the capabilities of quantum computing, particularly through the Variational Quantum Algorithm (VQA) framework, to manage complex high-dimensional problems effectively.

It is remarkable that a class of nonlinear optimal control problems could be formulated as a linear \Schrodinger \, wave equation, which could be solved using spectral methods aided by the quantum approach based on VQA. 
Subsequently, the quantum solution can be transformed back to obtain the classical solution of the optimal control problem in the limit as $h_w \to 0$.

The remainder of this paper is organized as follows.
\mref{sec:opt_control} provides a brief overview of optimal control theory. \mref{sec:generalized_hamiltonian} introduces the canonical quantization of constrained Hamiltonian systems, laying the theoretical groundwork for our subsequent exploration. We then proceed in \mref{Sec:Canonical_quantization} to apply these concepts to the quantization of optimal control problems. Specifically, \mref{Sec:oneDProb} focuses on one-dimensional problems, describing a spectral methods based approach for solving the \Schrodinger \, equation corresponding to optimal control problem formulation (\mref{Sec:spectraloneD}) and presenting an illustrative example (\mref{sec:analyticexample}). \mref{Sec:highdim} extends this analysis to higher-dimensional formulations. In \mref{sec:VQA}, we introduce Variational Quantum Algorithms (VQAs) as computational tools to address these complex problems, with \mref{subsec:VQA_non_Hermitian} highlighting the application of VQE (Variational Quantum Eigensolver) for non-Hermitian Systems. Computational examples, presented in \mref{sec:Simulations} (Examples 1 to 6), demonstrate the effectiveness of our proposed methodologies in practical settings. Finally, \mref{sec:conclusion} summarizes our conclusions.

\section{Optimal control problem}
\label{sec:opt_control}
 Consider a dynamical system described by the state equation:
\begin{equation}
\dot{\bm{x}}(t) = \bm{f}(\bm{x}(t), \bm{u}(t), t),
\end{equation}
where \( \bm{x}(t) \in \mathbb{R}^n \) is the state vector, \( \bm{u}(t) \in \mathbb{R}^r \) is the control input, and \( t \) is time.
The objective is to find a control function \( \bm{u}^*(t) \) that minimizes a given performance index \( J \), defined as:
\begin{equation} \label{eq:costfunction}
J = \Phi(\bm{x}(t_f)) + \int_{t_0}^{t_f} {L}(\bm{x}(t), \bm{u}(t), t) \, dt , 
\end{equation}
with the initial condition 
\begin{align}
    \bm{x}(t_0) = \bm{x}_0,
\end{align}
and assuming that $x(t_f)$ is free.
Here  \( \mathcal{L}(\bm{x}(t), \bm{u}(t), t) \) is the instantaneous cost or Lagrangian,
    \( \Phi(\bm{x}(t_f)) \) is the terminal cost, and
    \( t_0 \) and \( t_f \) represent the initial and final times, respectively.

The Pontryagin's Minimum Principle, introduced by the Russian mathematician Lev Pontryagin, provides necessary conditions for a control to be optimal and to solve the above problem. 
The Pontryagin Minimum Principle states that if $\bm{u}^*(t)$ is an optimal control and $\bm{x}^*(t)$ is the corresponding optimal trajectory, then there exists a non-zero adjoint variable (or costate variable) $\bm{\lambda}(t) \in \mathbb{R}^n$ and a Hamiltonian function $H: \mathbb{R}^n \times \mathbb{R}^r \times \mathbb{R}^n \times \mathbb{R} \to \mathbb{R}$ defined as:
\begin{align}
H(\bm{x}, \bm{u}, \bm{\lambda}, t) = \mathcal{L}(\bm{x}(t), \bm{u}(t), t)  + \bm{\lambda}^T \cdot \bm{f}(\bm{x}, \bm{u}, t),    
\end{align}
such that the following conditions hold:
\begin{align}
    \dot{\bm{\lambda}}^{*}(t) &= -\frac{\partial H}{\partial \bm{x}}(\bm{x}^*(t), \bm{u}^*(t), \bm{\lambda}^*(t), t), \\    
        \dot{\bm{x}}^{*}(t) &= \frac{\partial H}{\partial \bm{\lambda}}(\bm{x}^*(t), \bm{u}^*(t), \bm{\lambda}^*(t), t), \\
        \bm{x}^*(t_0) &= \bm{x}_0,
\end{align}
and the optimal solution $\bm{u}^*(t)$ satisfies 
\begin{equation}
    H(\bm{x}^*(t), \bm{u}^*(t), \bm{\lambda}^*(t)(t), t) \leq H(\bm{x}(t), \bm{u}(t), \bm{\lambda}(t), t), \quad \text{for all } \bm{u}(t) \in \mathbb{R}^m \text{ and all } t \in [t_0, t_f].
\end{equation}

These conditions are necessary for optimality, but not always sufficient. The Pontryagin Minimum Principle provides a powerful tool for deriving optimal control laws and solving optimal control problems.

\section{Canonical quantization of constrained Hamiltonian systems}
\label{sec:generalized_hamiltonian}

In classical mechanics, the Hamiltonian formulation provides a powerful framework for describing the dynamics of a system. 
Given the Hamiltonian, one can apply Dirac's method (c.f.~\cite{dirac2001lectures}) to obtain a first approximation to the corresponding quantum theory. Dirac's method is quite general and it can also deal with constrained systems, where certain relations among the coordinates and momenta are imposed. Indeed, Dirac brackets offer a systematic way to handle constraints.
In the following section, we will briefly review and outline the procedure for deriving a corresponding quantum theory from the given classical formulation. This section presents well-known results from Dirac's program for the canonical quantization of a classical physical system with constraints. The content is based on references \cite{dirac2001lectures, henneaux1992quantization, matschull1996dirac}, which interested readers can consult for more details.

The process of canonical quantization involves the transition from classical mechanics to quantum mechanics, wherein classical observables are systematically replaced by their quantum counterparts in the form of operators. In the following, we will briefly describe how this process is carried out for both unconstrained and constrained systems.

The classical (non-relativistic) motion of a physical system can be represented by a path $\bm{q}(t)$ in its \textit{configuration space} $\Xx$. We recall that \textit{configuration space} is the space (or manifold) of all possible positions or configurations the system can assume. For example, if a particle is restricted to move on a circle $\Ss^1$ in $\RR^2$ then its configuration space is the one-dimensional manifold $\Ss^1$.

The classical dynamics of a physical system is conventionally described through the Lagrangian, denoted as $L(\bm{q}, \dot{\bm{q}})$, where the generalized coordinate $\bm{q}$ represents a point in the configuration space $\Xx$ and $\dot{\bm{q}}$ its time derivative. 
If the system has $n$ degrees of freedom, then $\bm{q} = [q^1 \, q^2 \, \cdots q^n]^T $ and $\dot{\bm{q}} = [\dot{q}^1 \, \dot{q}^2  \, \cdots \dot{q}^n ]^T $. 

Considering an autonomous system, the Lagrangian $L(\bm{q}, \dot{\bm{q}})$ is defined as the difference between the system's kinetic energy $T$ and potential energy $V$, i.e., 
\begin{equation}
L(\bm{q}, \dot{\bm{q}}) = T - V.
\end{equation}
According to the \textit{action principle}, formulated by Hamilton, the true path taken by the particle between two points in its configuration space is the one for which the action integral is minimized or extremized. The action $\Ss$ is defined as the integral of the Lagrangian over time: 
\begin{align}
     \Ss = \int_{t_1}^{t_2} L \, dt,
\end{align}
and the extremization of the action integral takes place when its value is stationary under any variation $\delta \bm{q}(t)$ that vanishes at the endpoints, i.e., $ \delta \bm{q}(t_1) = \delta \bm{q}(t_2) = 0$.
The above condition implies that the path taken by the system (or the time evolution of the system) according to the action principle must satisfy the following Euler-Lagrange
equations 
\begin{equation}
    \frac{d}{dt} \left( \frac{\partial L}{\partial \dot{q}^i} \right) - \frac{\partial L}{\partial q^i} = 0.
\end{equation}
 For the Lagrangian $L(\bm{q}, \dot{\bm{q}})$, the corresponding Hamiltonian, $H(\bm{q}, {\bm{p}})$, is obtained through a Legendre transformation, where $\bm{p}$ represents the canonical momenta:
\begin{equation} \label{eq:defH}
H(\bm{q}, {\bm{p}}) =  p_i \dot{q}^i - L.
\end{equation}
Here,  $p_i = \frac{\partial L}{\partial \dot{q}^i} $ represents the conjugate momentum corresponding to $q^i$. Also, Einstein's summation notation was used in \meqref{eq:defH} and will be used in what follows. If the Jacobian matrix of the transformation  $(q^i,\dot{q}^i) \to (q^i,p_i) $, which is given by
\begin{equation}
    \pdir{p_i}{\dot{q}^j} = \frac{\partial^2 L }{\partial {\dot{q}^j} \partial\dot{q}^i},
\end{equation}
is invertible, then the Hamiltonian obtained by this transformation is unique. However, for a constrained physical system one should allow the possibility of a singular Jacobian matrix and consequently that of a nonunique Hamiltonian.
 We note that the space of $(\bm{q},\bm{p})$ is called the \textit{phase space} $\Pp$, which is indeed the cotangent bundle $T^* \Xx$ of the configuration space $\Xx$ (note that the momentum vector $\bm{p}$  is a linear functional on the tangent space $T_{\bm{q}} \Xx$). A point in the phase space $\Pp$ is called a \textit{state} and the time evolution of the system is a path in the phase space $\Pp$ that connects the initial and the final points.  
The canonical equations of motion  are given by:
\begin{equation} \label{eq:qdotpdot}
\dot{q}^i = \pdir{H}{p_i}, \quad \dot{p}_i = - \pdir{H}{q^i}.
\end{equation}
Alternatively, the equations of motion can be nicely expressed using Possion brackets:
\begin{equation}
\dot{q}^i = \{q^i, H\}, \quad \dot{p}_i = \{p_i, H\},
\end{equation}
where $\{ \,\, \cdot \,\,  ,\, \, \cdot \, \, \}$ represents the Poisson bracket.
We recall that for canonical coordinates $\bm{q}$ and momenta $\bm{p}$, the Poisson bracket of two functions $A(\bm{q}, \bm{p})$ and $B(\bm{q}, \bm{p})$ is defined as:
\begin{equation} \label{eq:Poissondef}
\{A, B\} =  \frac{\partial A}{\partial q^i} \frac{\partial B}{\partial p_i} - \frac{\partial A}{\partial p_i} \frac{\partial B}{\partial q^i}.
\end{equation}
Poisson brackets have the following properties:
\begin{enumerate}
    \item \textit{Linearity:} $\lcb{cA + B, C} = c\lcb{A, C} + \lcb{B, C}$.
    \item \textit{Leibniz (product) rule:} $\{AB, C\} = A\{B, C\} + \{A, C\}B$.
    \item \textit{Antisymmetry:} $\{A, B\} = -\{B, A\}$.
    \item \textit{Jacobi Identity:} $\{A, \{B, C\}\} + \{B, \{C, A\}\} + \{C, \{A, B\}\} = 0$.
\end{enumerate}
In the quantization scheme, classical observables, characterized by functions of generalized coordinates and momenta, find representation as Hermitian operators in the quantum domain. For a classical observable denoted by $A(\bm{q}, \bm{p})$, its corresponding quantum operator is expressed as $\hat{A} = A(\hat{\bm{q}}, \hat{\bm{p}})$. 
In the process of quantization, the classical Poisson bracket $\{A, B\}$ undergoes a transformation into the quantum commutator $[\hat{A}, \hat{B}]$, defined as:
\begin{equation}
[\hat{A}, \hat{B}] = \hat{A}\hat{B} - \hat{B}\hat{A}.
\end{equation}
More precisely, in the corresponding quantum theory, the classical Poisson bracket $\{A, B\}$ is transformed as follows: 
\begin{equation}
\{A, B\} \to \frac{1}{i \hbar} [\hat{A}, \hat{B}], 
\end{equation}
i.e., the classical Poisson bracket $\{A, B\}$ is replaced with $\frac{1}{i \hbar} [\hat{A}, \hat{B}]$, where $\hbar$ is the reduced Planck constant.
Corresponding to the classical Hamiltonian, which represents the total energy of a system,  the quantum Hamiltonian, denoted as $\hat{H}$, is obtained thought the quantization procedure described above. The quantum  Hamiltonian $\hat{H} = H(\hat{\bm{q}}, \hat{\bm{p}})$ determines the evolution of quantum states
via the time-dependent Schrödinger equation: 
\begin{equation}
   i\hbar \frac{\partial \Psi}{\partial t} = \hat{H} \Psi. 
\end{equation}
Here, $\Psi(\bm{q}, t)$ is the wave function.

While Poisson brackets serve as a powerful tool for expressing the fundamental principles of classical mechanics, they fall short when describing the dynamics of a system with constraints. For such cases, a more elegant approach exists, as introduced by Dirac through the use of so-called Dirac brackets. \\
\noindent \textbf{Overview of Dirac Brackets:}
Consider a system with canonical coordinates $\bm{q}$ and momenta $\bm{p}$, such that the primary constraints are given by 
\begin{equation} \label{eq:pconstraints}
    \phi_\alpha(\bm{q}, \bm{p}) = 0,
\end{equation}
where $\alpha$ labels the primary constraints. (later secondary constraints will be discussed) and we assume that $\alpha \in \mseqset{1}{m}$. The primary constraints $\phi_m$ are also assumed to satisfy certain regularity conditions (c.f., \cite{henneaux1992quantization}). We assume that the space defined by constraint conditions given in \meqref{eq:pconstraints} is a submanifold of the phase space $\Pp$ and this submanifold is referred to as the primary constraint surface. Later secondary constraints will also be discussed and the constraint surface defined by all the constraint conditions will be denoted by $\Ss$. The Hamiltonian is then modified to include these constraints:
\begin{equation}
H_{\text{T}} = H +  u^\alpha \phi_\alpha.
\end{equation}
Here, $u^\alpha$ are arbitrary coefficients (which could be functions of $\bm{q}$ and $\bm{p}$). 
Using the standard techniques of calculus of variations, 
(c.f., \cite{henneaux1992quantization})
it follows that
\begin{align} 
    \dot{q}^i &=  \pdir{{H}}{p_i} +    u^\alpha  \pdir{\phi_\alpha}{p_i}, \label{eq:HTqdot} \\
     \dot{p}_i &=  -\pdir{H}{q^i} -    u^\alpha  \pdir{   \phi_\alpha}{q^i} \label{eq:HTpdot}.
\end{align}

Using the above equation of motion and the Poisson bracket notation, for any given function $f$ of $\bm{p}$ and $\bm{q}$,  one can concisely express $\dot{f}$ as
\begin{align} \label{eq:deffdot}
    \dot{f} =  \pdir{f}{q^i} \dot{q}^i + \pdir{f}{p_i} \dot{p}_i = \{f, H\} +  u^\alpha \{f, \phi_{\alpha}\}.
\end{align}
Dirac observed that the above can be written even more elegantly as 
\begin{equation} \label{eq:fdotdirac}
     \dot{f} = \pbr{f}{H_T} = \pbr{f}{ H +  u^\alpha \phi_\alpha},
\end{equation}
if the expression on the right is carefully and appropriately interpreted. Since, in general, the coefficients $u^\alpha$ are not functions of $\bm{p}$ and $\bm{q}$, the definition of the Poisson bracket as given in \meqref{eq:Poissondef} does not apply to \meqref{eq:fdotdirac}. In fact, the Poisson bracket appearing in \meqref{eq:fdotdirac} should be interpreted as follows. One can use the properties of the Poisson bracket to write
\begin{align} \label{eq:fdotdiracexplained}
   \pbr{f}{H_T}   
   & = \{f, H \} + \{ f,  u^\alpha\phi_{\alpha}\} \nonumber\\
     &= \{f, H \} +  \{ f,  u^\alpha\} \phi_{\alpha} +  u^\alpha \{ f, \phi_{\alpha} \} \nonumber\\
     & = \{f, H \} +  u^\alpha 
     \pbr{f}{\phi_{\alpha}},
\end{align}
where the constraint condition $\phi_{\alpha} =0 $ is used to obtain the last expression.
Clearly, the expressions of  $\dot{f}$ given in \meqref{eq:fdotdirac} and \meqref{eq:fdotdiracexplained}
match. Here, in interpreting the Poisson bracket $\pbr{f}{ H +  u^\alpha \phi_\alpha}$, the constraint condition $\phi_{\alpha} =0 $ was used at the end after expanding the Poisson bracket, otherwise one would obtain erroneous results. Dirac used the symbol ``$\approx$" to denote this and called this ``weak equality.'' In other words, unlike the usual notion of ``strong equality" the ``weak equality''  relations do not hold on the entire phase space $\Pp$, but only on the constrained surface defined by the all the constraint conditions  $\phi_{\alpha} =0 $, where $\alpha$ runs through the elements of the set indexing all the constraints. We note that, in general, the Poisson brackets are not compatible with the weak equalities (it means, $f \approx 0$ does not imply that $\pbr{f}{g} \approx 0$ for a general $g$). We further observe that, in the evaluation of the Poisson bracket in \meqref{eq:fdotdiracexplained}, the constraint conditions $\phi_{\alpha} =0 $ are used at the end (from going to the third step from the second step) after working out the expansion of the Poisson bracket in the second step.  This care must always be taken in evaluating the Poisson bracket that involves weak equality. Following this convention, it is clear that the expressions of  $\dot{f}$ given in \meqref{eq:fdotdirac} should more precisely be written as 
\begin{equation} \label{eq:fdotdiracweako}
     \dot{f} \approx \pbr{f}{H_T}.
\end{equation}
Next, we classify constraints as first-class and second-class constraints as follows.   \\
 \noindent \textbf{First-class constraints:} The constraint (\(\eta_{\gamma}\)) is called a first-class constraint, if its Poisson brackets with all other constraints (\(\phi_\beta\)) vanish weakly, i.e., \(\ \pbr{\eta_{\gamma}} {\phi_{\beta}} \approx 0\) for all constraints (\(\phi_\beta\)). The implication of first-class constraints is that they generate gauge transformations, preserving symmetries in the system without affecting its physical degrees of freedom. 
Using the properties of the Poisson brackets, including Jacobi identities, it can be shown that the Poisson brackets of two first-class constraints must also be first-class. \\
\noindent \textbf{Second-class constraints:} A constraint (\(\phi_\alpha\)) is called a second-class constraint, if there exists at least one constraint (\(\phi_\beta\)) such that \(\{\phi_\alpha, \phi_\beta\} \napprox 0\). 

Similar to the above classification, one can also characterize an arbitrary function $f(\bm{q},\bm{p})$  as first-class if \(\pbr{f} {\phi_{\beta}} \approx 0\) for all constraints (\(\phi_\beta\)), otherwise it is said to be of the second-class. 

Since constraint condition $\phi_{\beta} = 0$ implies that  $\dot{\phi}_{\beta} \approx 0$ (this is a fundamental consistency requirement, as the primary constraint conditions should remain preserved independent of time), on replacing $f$ by $\phi_{\beta}$ in \meqref{eq:fdotdiracweako}, for $\beta = \mseqenum{1}{m}$, the following equations are obtained as natural consistency conditions resulting from the primary constraints
\begin{equation} \label{eq:fdotdiracweak}
     \dot{\phi}_{\beta} \approx \pbr{\phi_{\beta}}{H_T} \implies  \{\phi_{\beta}, H \} +  
     \pbr{\phi_{\beta}}{\phi_{\alpha}} u^\alpha  \approx 0.
\end{equation}
For the moment, assume that we are dealing only with second-class constraints. 
Let $M = [ M_{\alpha\beta}]$ be an $m\times m$ matrix with matrix elements given by
\begin{align}\label{eq:defM}
    M_{\alpha\beta} = \pbr{\phi_{\alpha}}{\phi_{\beta}}.
\end{align}
Assuming matrix $M$ to be invertible, let us denote its inverse  as $M^{-1} = [M^{\alpha \beta}]$, where
\begin{equation}
    M^{\alpha \beta} M_{\beta \gamma} = \delta^{\alpha}_{\gamma}.
\end{equation}
Then, the above system consisting of $m$ linear equations has a unique solution, $u^\alpha$, and it can be computed as 
\begin{align} \label{eq:defualpha}
    u^\alpha \approx - M^{\alpha \beta} \pbr{\phi_\beta}{H}.
\end{align}
Then, from \meqref{eq:deffdot} and  \meqref{eq:defualpha} it follows that
\begin{equation} \label{eq:fdotdiracbracketdef}
     \dot{f}  \approx \pbr{f}{H} - 
     \pbr{f}{\phi_{\alpha}} M^{\alpha \beta} \pbr{\phi_\beta}{H}.
\end{equation}
Dirac proved that if one restricts to only the set of second-class constraints (also no linear combination of constraints in this set should be first-class), then the corresponding matrix $M$ is invertible. The above discussion motivates the following definition of the Dirac bracket.
\begin{defn} \label{def:DB}
    The Dirac bracket between two functions $A(\bm{q}, \bm{p})$ and $B(\bm{q}, \bm{p})$ is defined as:
\begin{equation}
    \label{eq:dirac_bracket_definition}
    \{A, B\}_{DB} = \{A, B\} - \{A, \phi_\alpha\} M^{\alpha \beta} \{\phi_\beta, B\},
\end{equation}
where $M^{-1} = [M^{\alpha \beta}]$ is the inverse of the matrix $M = [M_{\alpha \beta}]$ whose elements are Poisson brackets between the second-class constraint functions, i.e., $[M_{\alpha \beta}] = [\{\phi_\alpha, \phi_\beta\}]$. Furthermore, the summation is carried out over the set of all the second-class constraint functions such that no linear combination of constraints in this set results in a first-class constraint. 
\end{defn}
It can be verified that similar to Poisson brackets, Dirac brackets are also bilinear and antisymmetric, and satisfy the Leibniz rule and the Jacobi identity. It is also clear that for an unconstrained system, Dirac brackets reduce to the corresponding Poisson brackets.

We note that in the discussion above, we assumed that only second-class constraints were present and that the system of equations in  \meqref{eq:fdotdiracweak} has a solution. However, this may not be true in general and one may encounter other possibilities. For example, the system of equations in  \meqref{eq:fdotdiracweak} may be \textit{inconsistent}. It may indicate a problem with the formulation of the Lagrangian or the underlying physical model. For example, if the Lagrangian is of the form $L=q$, it results in equations of motion that are inherently inconsistent. The second possibility is that of the system of equations in  \meqref{eq:fdotdiracweak} being an identity (such as the equation of the form $1=1$). In this case, only the primary constraints are sufficient and the system of equations does not provide any new information. The third possibility is that  the system of equations in  \meqref{eq:fdotdiracweak} may lead to $k$ additional  constraints
\begin{equation}
    \phi_{\gamma}(\bm{q}, \bm{p}) = 0, 
\end{equation}
for $\gamma \in  \{m+1,\, m+2,\, \cdots,\, m+k \}$.
These constraints, which are not primary constraints but result from applying the consistency conditions to the primary constraints, are called  \textit{secondary constraints}. These secondary constraints are added to the extend the Hamiltonian and they lead to further consistency conditions (similar to \meqref{eq:fdotdiracweak}) given as   
\begin{equation}
 \dot{\phi}_{\gamma} \approx  \pbr{\phi_{\beta}}{H_T} \approx 0,   
\end{equation}
for $\gamma =  m+1,\, m+2,\, \cdots,\, m+k $. This system of equations can be treated as earlier, and it can result in various possibilities, as discussed previously. If it results in further secondary constraints, then the corresponding consistency conditions must be considered.  The process may be continued until no more secondary constraints are obtained and all the consistency conditions are exhausted. At this point, all the primary and secondary constraints are added to the extended Hamiltonian (in a way, the primary and the secondary constraints are now on the same footing), and the resulting Hamiltonian determines the time evolution of the physical system.

Assume at this point,  we have a total $N_f + N_s$ constraints which are classified into first-class and second-class constraints as follows:
\begin{equation}
     {\eta}_{\gamma}  = 0, \quad \text{and} \quad  {\phi}_{\alpha}  = 0, 
\end{equation}
where for $\gamma = 1$ to $\gamma = N_f$ are the first-class constraints and 
and 
for $\alpha = 1$ to $\alpha = N_s$ are the second-class constraints.
Then, we have the following expression for the total Hamiltonian $H_T$
\begin{align} \label{def:Htotal}
    H_T &= H + u^{\alpha} \phi_{\alpha} + v^{\gamma} \eta_{\gamma} 
     = H - M^{\alpha \beta} \pbr{\phi_\beta}{H} \phi_{\alpha} + v^{\gamma} \eta_{\gamma} 
     = H^{\prime} + v^{\gamma} \eta_{\gamma},
\end{align}
where 
\begin{equation} \label{eq:Hprime}
    H^{\prime} = H - M^{\alpha \beta} \pbr{\phi_\beta}{H} \phi_{\alpha},
\end{equation}
and $[M^{\alpha \beta}]$ is the inverse of an $N_s \times N_s$ matrix $M = [M_{\alpha \beta}]$ with matrix elements  $ M_{\alpha\beta} = \pbr{\phi_{\alpha}}{\phi_{\beta}}.
$

It can be checked that the Poisson bracket of any two first-class primary constraints generates a gauge transformation that does not affect the physical state of the system (see \cite{henneaux1992quantization}). 
According to Dirac's conjecture, first-class secondary constraints also generate gauge transformations. Under certain regularity assumptions, a proof of this conjecture was presented in \cite{henneaux1992quantization}. We will assume that all
first-class constraints generate gauge transformations. Therefore, the terms $v^{\gamma} \eta_{\gamma}$ corresponding to the first-class constraints in \meqref{def:Htotal} can be dropped from the Hamiltonian.
For any arbitrary function, $g (\bm{q}, \bm{p})$,  one can easily verify the following (see \meqref{eq:fdotdiracweako}):
\begin{align}
    \dot{g} \approx \pbr{g}{H_T} \approx \pbr{g}{H^{\prime}} \approx \dbr{g}{H}.
\end{align}
It is clear from the definition of the Dirac bracket (see Definition \ref{def:DB}) that
for a first-class function $\phi_f$, and any arbitrary function  $g(\bm{q},\bm{p})$, their Dirac bracket reduces to their Poisson bracket
\begin{equation} \label{eq:firstclasspbdb}
    \dbr{\phi_f}{g} \approx  \pbr{\phi_f}{g}.
\end{equation}
We recall that the weak equality ``$\approx$" is guaranteed to hold only on the constraint surface $\Ss$. It can be shown that if 
$f \approx g$, then 
\begin{equation}
    f - g = c^{\alpha} \phi_{\alpha},
\end{equation}
 where $c^{\alpha}$ are some functions of $\bm{q}$ and $\bm{p}$
(see Theorem 1.1, \cite{henneaux1992quantization}). It implies that if $f(\bm{q},\bm{p})$ is first-class, then 
\begin{equation} \label{eq:fexpansion}
    \pbr{f}{\phi_\alpha} = f_\alpha^\beta \, \phi_\beta,
\end{equation}
where $f_\alpha^\beta$ are functions of $\bm{q}$ and $\bm{p}$.

The process of quantization involves transforming the classical Dirac bracket $\dbr{A}{B}$ into the quantum commutator $[\hat{A}, \hat{B}]$ as follows
\begin{equation} \label{eq:dbrquantization}
\dbr{A}{B} \to \frac{1}{i \hbar} [\hat{A}, \hat{B}].
\end{equation}
Then the quantum  Hamiltonian $\hat{H}^{\prime} $ (corresponding to the classical Hamiltonian $H^{\prime}$ defined in \meqref{eq:Hprime}) governs the evolution of quantum system
via the time-dependent \Schrodinger \, equation: 
\begin{equation} \label{eq:defschrod}
   i\hbar \frac{\partial \Psi(x, t)}{\partial t} = \hat{H}^{\prime} \,  \Psi(x, t).
\end{equation}

It is important to note that there are some subtleties involved in the above quantization process using Dirac brackets and it was nicely explained by Dirac in the reference \cite{dirac2001lectures}. 
For example, consider the classical system with only first-class constraints, say $\eta_{\gamma}$ for $\gamma \in \mseqset{1}{N_f} $. For the quantization process, Dirac postulated that the wave function $\Psi$ should be annihilated by the operators 
 corresponding to constraints:
\begin{equation} \label{eq:firstclasskillpsi}
    \hat{\eta}_{\gamma} \Psi = 0, \quad \gamma \in \mseqset{1}{N_f}.
\end{equation}
 It means, one also must have
\begin{equation}
    [\hat{\eta}_{j},\hat{\eta}_{k}]  \Psi = 0, \quad \text{for any } j,k \in \mseqset{1}{N_f}.
\end{equation}
Since in the classical theory, $\pbr{\eta_j}{\eta_k} \approx 0$,  it follows from \meqref{eq:fexpansion} that
\begin{equation}
    \pbr{\eta_j}{\eta_k} = c_{jk}^{\alpha} \eta_{\alpha}.
\end{equation}
Therefore, in the quantum theory the following consistency conditions must be satisfied:
\begin{equation}
    \lsb{\hat{\eta}_j,{\hat{\eta}_k}} = \hat{c}_{jk}^{\alpha} \hat{\eta}_{\alpha}.
\end{equation}
In addition, corresponding to the classical consistency  condition $\dot{\eta}_{\gamma} \approx 0$ or equivalently $ \pbr{\eta_{\gamma}}{H} \approx 0$, one should have the quantum relation 
\begin{equation}
     \lsb{\hat{\eta}_{\gamma}, \hat{H}} \, \Psi = 0, \quad 
     \gamma \in \mseqset{1}{N_f}.
\end{equation}
As before, this leads to the following consistency conditions:
\begin{equation}\label{eq:firstclasskillpsiH}
    \pbr{\hat{\eta}_{\gamma}} {\hat{H}} = \hat{d}_{\gamma}^{\alpha} \hat{\eta}_{\alpha}.
\end{equation}
Since $\hat{c}_{jk}^{\alpha}$ and $\hat{d}_{\gamma}^{\alpha}$ are in general functions of $\hat{\bm{q}}$
and $\hat{\bm{p}}$, they will not commute with $\hat{\eta}_{\alpha}$, therefore the order of operators is important. In order to get a satisfactory quantum theory, these coefficients $\hat{c}_{jk}^{\alpha}$ and $\hat{d}_{\gamma}^{\alpha}$ must be on the left in the above equation. When this is not possible, one cannot achieve an accurate quantum theory.

We note that if only the first-class constraints are present, as we assumed above, then only the Poisson brackets are sufficient to get the quantum theory, as discussed above.
However, if second-class constraints are present, then in transitioning to the quantum theory, Dirac brackets are needed, and one must use \meqref{eq:dbrquantization} for quantization. For example, consider the constraints
\begin{align}
    q^1 =  0 \quad \text{and} \quad p_1 = 0.
\end{align}
If one naively obtains $\cbr{\hat{q}^1}{\hat{p}_1}$ via the usual canonical quantization, i.e., by using Poisson bracket and replacing $ \pbr{q^1}{p_1}$ by $ \frac{1}{i \hbar}\cbr{\hat{q}^1}{\hat{p}_1}$, then it leads to inconsistencies as shown below:
\begin{align}
     \hat{q}^1 {\Psi} = 0,  \quad \hat{p}_1 {\Psi} = 0 \implies  \cbr{\hat{q}^1}{\hat{p}_1} \, {\Psi} = 0 
     \implies i \hbar \Psi = 0.
\end{align}
We note that even though $\pbr{q^1}{p_1} =1 \neq 0$, we have $\dbr{q^1}{p_1} = 0$. Therefore, the use of Dirac bracket does not lead to the above inconsistency. 
Furthermore, we also observe that for a second-class constraint $\phi_s$,  we have
\begin{align}\label{eq:gsec}
     \dbr{g}{\phi_s} = \pbr{g}{\phi_s} - \pbr{g} {\phi_\alpha} M^{\alpha \beta} \pbr{\phi_\beta} {\phi_s} = 0.
\end{align}
Therefore, the secondary constraints can be considered $ \phi_s = 0$, before working out the Dirac brackets and it holds strongly, i.e.,  
\begin{align} \label{eq:secondstrong}
    \phi_s = 0
\end{align}
on the entire phase space.

To summarize, in transitioning to quantum theory the Dirac brackets are replaced by the commutator relations as given in \meqref{eq:dbrquantization}, the second class constraints are treated as strongly equal to $0$ (see \meqref{eq:secondstrong}), and the first class constraints are postulated to annihilate the wave function and the resulting conditions as given in \meqref{eq:firstclasskillpsi} to \meqref{eq:firstclasskillpsiH}. Further, if only second-class constraints are present, then  one can work with the original Hamiltonian $H$  instead of $H^{\prime}$ or $H_T$ \cite{rothe2010classical} and the \Schrodinger \, equation becomes 
\begin{equation} \label{eq:defschrodshort}
   i\hbar \frac{\partial \Psi(x, t)}{\partial t} = \hat{H} \,  \Psi(x, t).
\end{equation}

The following example provides an illustration of some of the concepts discussed above that are relevant to the canonical quantization of a constrained Hamiltonian system. 
\begin{example}
    Let us consider the system with Hamiltonian $H$ 
     with the primary constraints $\phi_1$ and $\phi_2$ given as 
    \begin{align}
        H = \frac{1}{2} {x}^2 e^{y}, \quad \phi_1 \equiv p_x = 0, \quad \phi_2 \equiv p_y = 0.
    \end{align}
    We have the total Hamiltonian  $H_T = H + u^{1} \phi_1 + u^{1} \phi_2$. 
    From the primary constraints the consistency requirements $\dot{\phi}_1 = \dot{p}_x \approx 0$ and 
    $\dot{\phi}_2 = \dot{p}_y \approx 0$ lead to the following secondary constraint: 
    \begin{align}
        \dot{p}_x & \approx 0 \implies \pbr{p_x}{H_T} \approx 0 \implies - \pdir{H}{x} \approx 0 \implies x \approx 0, \\ 
         \dot{p}_y & \approx 0 \implies \pbr{p_y}{H_T} \approx 0 \implies - \pdir{H}{y} \approx 0 \implies x \approx 0.
    \end{align}
    Therefore, we obtain $\phi_3 \equiv x = 0$ as a secondary constraint. Further, consistency requirement $\dot{\phi}_3 = \dot{x} \approx 0 $ does not lead to any new constraints. At this stage, we have two primary constraints ($\phi_1 =0 $ and $\phi_2 =0 $) and one secondary constraint ($\phi_3 = 0$). It can be verified that,
    \begin{align}
        \pbr{\phi_1}{\phi_2} =0, \quad \pbr{\phi_1}{\phi_3} = -1, \quad  \pbr{\phi_2}{\phi_3} = 0. 
    \end{align}
    It follows that ${\phi_2} = 0$ is a first-class constraint and ${\phi_1} = 0$, ${\phi_3} = 0$ are second-class constraints. Using only the second-class constraints, the matrix $M$ (ref.~ \meqref{eq:defM}) can easily be computed as follows.
    \begin{align}
        M = \bmat{\pbr{\phi_1}{\phi_1}}{\pbr{\phi_1}{\phi_3}}{\pbr{\phi_3}{\phi_1}}{\pbr{\phi_3}{\phi_3}} = \mat{0}{-1}{1}{0}.
    \end{align}
    Clearly, the matrix $M$ is invertible as it should be. 
    Next, using the definition of Dirac bracket (Def.~\ref{def:DB}), one can easily compute
    \begin{align}
       \dbr{x}{y} = 0, \quad  \dbr{x}{p_x} = 0, \quad \dbr{x}{p_y} = 0, \quad  \dbr{y}{p_x} = 0, \quad \dbr{p_x}{p_y} = 0, \quad \text{and} \quad  \dbr{y}{p_y} = 1. 
    \end{align}
Finally, for quantization, the classical Dirac brackets are transformed  into the quantum commutator as follows
\begin{align}
      [\hat{x}, \hat{y}] = 0, \quad  [\hat{x}, \hat{p}_x] = 0, \quad [\hat{x}, \hat{p}_y] = 0 , \quad [\hat{y}, \hat{p}_x] = 0, \quad [\hat{p}_x, \hat{p}_y] = 0 \quad \text{and} \quad  [\hat{y}, \hat{p}_y] =  i \hbar. 
    \end{align}
Next, the total Hamiltonian at this stage is expressed as 
\begin{equation}
    H_T = H + u^{1} \phi_1 + u^{2} \phi_2 + u^{3} \phi_3, \end{equation}
where,
\begin{equation*}
    \twobyone{u^{1}}{u^{3}} =  - M^{-1}  \twobyone{\pbr{\phi_1}{H}}{\pbr{\phi_3}{H}} = \twobyone{-\pbr{\phi_3}{H}}{\pbr{\phi_1}{H}} =  \twobyone{0}{-x e^y}. 
\end{equation*}
On further simplification one obtains 
    \begin{equation}
        H_T = -\frac{1}{2} x^2 e^y + u^2 \phi_2 =  H^\prime + u^2 \phi_2, 
    \end{equation}
    where $H^\prime = -\frac{1}{2} x^2 e^y$. 
The time evolution of the quantum system is determined by the \Schrodinger \,  equation \meqref{eq:defschrod}.

\end{example}

\section{Canonical quantization of optimal control problems} \label{Sec:Canonical_quantization}

Based on Dirac's canonical quantization theory discussed in the previous section, in this section we present the quantization of optimal control problems. In \mref{Sec:oneDProb}, we describe the canonical quantization of a class of one-dimensional problems, followed by a description of a spectral methods-based approach for solving the \Schrodinger \, equation corresponding to the optimal control problem formulation given in \mref{Sec:spectraloneD}. An illustrative example is presented in \mref{sec:analyticexample}. In \mref{Sec:highdim}, this analysis is extended to multi-dimensional optimal control problems.

\subsection{One dimensional problems}
\label{Sec:oneDProb}
Consider the system described by the Lagrangian
\begin{align}
   L & = L(x(t), u(t)), 
\end{align}
with the constraint 
\begin{equation} \label{eq:xdotDef}
    \dot{x}(t) = f(x(t), u(t)),
\end{equation}
where $x(t_0) = x_0$.
Our objective is to minimize the following performance index (PI).
\begin{equation} \label{eq:costfunctionnew}
J = \Phi({x}(t_f)) + \int_{t_0}^{t_f} {L}({x}(t), {u}(t)) \, dt. 
\end{equation}
We form the Hamiltonian using the Legendre transform
\begin{align}
    H = \dot{x} p_x + \dot{u} p_u -L,
\end{align}
with the constraints:
\begin{align}
       \phi_1 &\equiv p_u  = 0, 
\end{align}
and $ \dot{x} =  \pdir{H}{p_x} = f$.
This leads to the following Hamiltonian.
\begin{align}
    H_T = -L + f p_x + v\textsuperscript{1} \phi_1,
\end{align}
where $v^1$ is an arbitrary function.
Clearly, for this Hamiltonian $\dot{x} = \pdir{H}{p_T} = f$ is satisfied throughout the evolution of the system dynamics.
Next, the consistency condition requires that $ \dot{\phi}_1 \approx 0$. This leads to the following additional constraint:
\begin{align}
     \dot{\phi}_1  \approx 0 \implies \pbr{p_u}{H_T} \approx 0 \implies - \pdir{H_T}{u} \approx 0 \implies \pdir{\lob{L - f p_x}}{u} \approx 0.
\end{align}
Let us define \begin{align}
    \theta = L - f p_x .
\end{align}
The new constraint can be expressed as 
\begin{equation}
    \phi_2 \equiv \theta_u = 0.
\end{equation}
Therefore, the extended Hamiltonian now is 
\begin{align} \label{eq:extendenHT}
     H_T = -L + f p_x + v^1 \phi_1 + v^2 \phi_2 = -\theta + v^1 p_u + v^2 \theta_u. 
\end{align}
The consistency conditions demand $\dot{\phi}_1 \approx 0$ and $ \dot{\phi}_2 \approx 0$. 
These conditions do not lead to new constraints and can be satisfied by an appropriate choice 
of $v^1$ and $v^2$ that solves the following equations:
\begin{align}
    \dot{p}_u & \approx \pbr{p_u}{H_T} = \theta_u - v^2 \theta_{uu}  = 0,   \\
    \dot{\theta}_u & \approx \pbr{\theta_u}{H_T}  = f \theta_{ux}  - f_u \theta_x  + v^1 \theta_{uu}  = 0,
\end{align}
where we assumed that $\theta_{uu} \neq 0$. The solutions for the above equations are given by 
\begin{align} \label{eq:v2v1}
    v^2 = \frac{\theta_u}{\theta_{uu}}, \quad \text{and } v^1 = 
    \frac{\lob{f_u \theta_x - f \theta_{ux}}}{\theta_{uu}}.
\end{align}
It is easy to check that 
\begin{align}
    \pbr{\phi_1}{\phi_2} =   \pbr{p_u}{\theta_u} = - \theta_{uu}.
 \end{align}
 Therefore,
 \begin{align} \label{eq:M}
        M = \twobytwo{\pbr{\phi_1}{\phi_1}}{\pbr{\phi_1}{\phi_2}}{\pbr{\phi_2}{\phi_1}}{\pbr{\phi_2}{\phi_2}} = \twobytwo{0}{-\theta_{uu}}{\theta_{uu}}{0}.
\end{align}
Also, $M$ is invertible if  $\theta_{uu} \neq 0$ as we have assumed.
Then, 
\begin{align} \label{eq:MInv}
        M^{-1} =  \frac{1}{\theta^2_{uu}} \twobytwo{0}{\theta_{uu}}{-\theta_{uu}}{0}.
\end{align}
We note that \meqref{eq:v2v1} can also be obtained using the general theory (i.e., from \meqref{eq:defualpha}) as 
\begin{equation*}
    \twobyone{v^1}{v^2} =  - M^{-1}  \twobyone{\pbr{\phi_1}{H}}{\pbr{\phi_2}{H}} = \frac{1}{\theta_{uu}}\twobyone{f_u \theta_x - f \theta_{ux}}{\theta_u}.
\end{equation*}
Here, both the constraints are second class therefore from the general theory, as noted previously (see \meqref{eq:gsec}), their Dirac brackets with any arbitrary function $g$ vanishes, i.e.,  
$\dbr{g}{\phi_1} = \dbr{g}{\phi_2} = 0$. 
Next, we compute,
\begin{align}
  \dbr{x}{p_x} & = 1, \\
    \dbr{x}{u} &= \frac{f_u}{\theta_{uu}}, \\
    \dbr{p_x}{u} &=  \frac{\theta_{ux}}{\theta_{uu}}. 
\end{align}
Next, using the standard quantization procedure given by Dirac, we consider the operators $\hat{x}$, $\hat{p}_x$, $\hat{u}$ and $\hat{p}_u$ associated with $x$, $p_x$, $u$ and $p_u$, respectively. Before proceeding further to obtain
the corresponding commutator relations, we define a symmetrization operator ``$\mwidetilde{(\,\cdot\,)}$'' for arbitrary  operators $\hat{g}_1$ and  $\hat{g}_2$ as 
$$\mwidetilde{\hat{g}_1 \hat{g}_2} = \frac{1}{2} \lob{\hat{g}_1 \hat{g}_2 + \hat{g}_2 \hat{g}_1}.
$$
Further, let us assume that  $h_r$ represents a parameter that plays the role of Planck's constant $\hbar$
in our theory.
The commutator relations are: 
\begin{align}
\lsb{\hat{x}, \hat{p}_x } & = i h_r,  \\
    \lsb{\hat{x}, \hat{u}} &= i h_r \mwidetilde{\frac{{f}_u} {\theta_{uu}}}, \label{eq:commutatorxu} \\
    \lsb{\hat{p}_x, \hat{u}} &= i h_r \mwidetilde{\frac{\theta_{ux}}{\theta_{uu}} }. \label{eq:commutatorpxu}
\end{align}
It easily follows that 
\begin{align}
    \hat{p}_x  = \frac{h_r}{i} \pdir{}{x}.
\end{align}
Next, we determine the corresponding differential operator for $\hat{u}$. Let us express \begin{align*}
\hat{u} &= \mathcal{M}(\hat{x}) \pdir{}{x} + \mathcal{N}(\hat{x}),
\end{align*} 
where $\mathcal{M}(x)$ and $\mathcal{N}(x)$ are arbitrary functions of $x$. 
We also assume that $\theta_{uu}$ is a constant.
Then it follows from an easy calculation that, 
\begin{align} \label{eq:MN}
    \mathcal{M} &= - i h_r \frac{f_u}{\theta_{uu}} \quad \text{and} \quad  \mathcal{N}' =  - \lob{ \mwidetilde{\frac{\theta_{ux}}{\theta_{uu}}}  + \mathcal{M}' \pdir{}{x}}.
\end{align}

Next, we consider the following special case.
\begin{align} 
    L(x,u) &= \frac{m u^2}{2} + V(x) \label{eq:defL}\\
     \dot{x} &= f(x,u) = a(x) +  b(x) u . \label{eq:defxdot}
\end{align}
 Then we have,
\begin{align}
  \theta = \frac{mu^2}{2} + V(x) - \lob{a(x) +  b(x) u } p_x, \quad  \theta_{uu} = m, \quad 
    f_u = b(x)
\end{align}
and the following symmetrization for $\theta_{ux}$,
\begin{align}
   \mwidetilde {\theta_{ux}} = - \frac{1}{2} \lob{b^\prime(\hat{x}) \,  \hat{p}_x  + \hat{p}_x \,  b^\prime(\hat{x})} =   \frac{i h_r}{2} \lob{b^\prime(\hat{x}) \,  \pdir{}{x}  + \pdir{}{x}  \,  b^\prime(\hat{x})}. 
\end{align}
Then, by using \meqref{eq:MN}, $\mathcal{M}$ and $\mathcal{N}$ can be computed by applying the commutator relations given in \meqref{eq:commutatorxu} and \meqref{eq:commutatorpxu} on a test function. This leads to the following expression for the $\hat{u}$,
\begin{align} \label{eq:defcontrolOP}
    \hat{u} = - \frac{i h_r}{m} \lob{ b(\hat{x}) \pdir{}{x} + \frac{b^\prime (\hat{x})}{2 }}.
\end{align}
Therefore, we obtain the Hamiltonian (using an appropriate symmetrization of operators involved) as
\begin{align}
    \hat{H} &= - \frac{m \hat{u}^2}{2} - V(\hat{x}) + 
    \lob{b(\hat{x}) \hat{u} + a(\hat{x})} \hat{p}_{x} \\
    &= - \frac{h_r^2}{2m} \lob{b(\hat{x}) \pdir{}{x} + \frac{b^\prime (\hat{x})}{2 }}^2 -  V(\hat{x}) - i h_r \lob{ a(\hat{x}) \pdir{}{x} + \frac{a^\prime(\hat{x})}{2 }}.
\end{align} 
The corresponding  \Schrodinger \, equation is given by 
\begin{equation} \label{eq:defschrodoneD}
   i h_r \frac{\partial \Psi(x, t)}{\partial t} = \hat{H} \,  \Psi(x, t).
\end{equation}
Assuming that, 
\begin{equation}
\label{eq:defRphi}
   \Psi(x, t) = R(x,t) \exp \lob{\frac{i}{h_r} S(x,t)},  
\end{equation}
it can be verified that the $S(x,t)$ satisfies the following generalized Hamilton-Jacobi equation
\begin{equation}
    \pdir{S}{t} + \frac{b^2}{2m} \lob{\pdir{S(x,t)}{x}}^2 + a \pdir{S(x,t)}{x} - V(x) - V_a(x) = 0,
\end{equation}
where $V_a(x)$ is the additional cost term defined as 
\begin{equation}
    V_a(x) = \frac{h_r^2}{2m}
    \left\{ \left( \frac{1}{2} b \frac{d^2 b}{dx^2} + \frac{1}{4} \left( \frac{db}{dx} \right)^2 \right) +  \frac{1}{R} \lob{2b \frac{db}{dx} \frac{\partial R}{\partial x}  + b^2 \frac{\partial^2 R}{\partial x^2}} \right\}. 
\end{equation}
We also note that a terminal condition \( S(x, t_f) = - \Phi(x) \) at \( t = t_f \) results in the terminal wave function as 
\begin{equation} \label{eq:terminal}
    \Psi(x, t_f) = R (x, t_f) \exp(-\frac{i}{h_r}  \Phi(x)),
\end{equation}
where we note that $\Phi({x}(t_f))$ is the terminal cost in \eqref{eq:costfunctionnew}.
Finally, using \meqref{eq:defcontrolOP} and \meqref{eq:defRphi} the control $u(x,t)$ can be obtained as 
\begin{align}
    u(x,t) = \frac{b}{m} \pdir{S(x,t)}{x} - \frac{i h_r}{m} \lob{ \frac{b}{R} \pdir{R}{x} + \frac{db}{dx}}.
\end{align}

Next comes a crucial step. Here we transform the \Schrodinger \, equation \eqref{eq:defschrodoneD} by substituting 
\( h_r = i {h}_w \) where \( {h}_w \) is real, and transforming the Hamiltonian operator as
\[
\hat{H}_w = -\hat{H} \Big|_{h_r = i {h}_w}.
\]
This results in the modified \Schrodinger \, equation given by 
\begin{equation} \label{eq:defschrodoneDw}
    h_w \frac{\partial \Psi(x, t)}{\partial t} = \hat{H}_w \,  \Psi(x, t).
\end{equation}
Assuming that 
$R$ is chosen so that $\lob{ \frac{b}{R} \pdir{R}{x} + \frac{db}{dx}}$ remains bounded, as $h_w \to 0$,
we get
\begin{align} \label{eq:uhat}
    u(x,t) = \frac{b(x)}{m} \pdir{S(x,t)}{x}.
\end{align}
One can obtain the optimal control variable $ u^*(x,t)$ using \meqref{eq:defRphi} as
\begin{align} \label{eq:defufinal}
    u^*(x,t)  = \frac{b(x)}{m} \lob{\frac{h_w \, R(x,t) }{\Psi(x,t)}}  \, \pdir{}{x} \lob{\frac{\Psi(x,t)}{R(x,t)}}.
\end{align}
Once the optimal control variable $ u^*(x,t)$ is known, the optimal state variable $ x^*(x,t)$ can be obtained by solving the differential equation \meqref{eq:xdotDef}.

\subsubsection{Spectral method for solution of the Schr\"{o}dinger equation corresponding to the optimal control problem} \label{Sec:spectraloneD}
Next, we describe a spectral method for obtaining the solution of the \Schrodinger \, equation \meqref{eq:defschrodoneDw} corresponding to the optimal control problem described earlier. We approximate the wave function $\Psi(x, t)$ using  orthonormal basis functions \( \psi_n(x) \) of a state variable \( x \):
\begin{align} \label{eq:orthonormalexpansion}
    \Psi(x, t) = \sum_{n=1}^{N} c_n(t) \psi_n(x).
\end{align}
Suitable orthonormal basis functions \( \psi_n(x) \) for a state variable \( x \) may include sine, cosine, Chebyshev, Legendre, or Jacobi basis functions. 
Also, the number of terms $N$ in the expansion in \meqref{eq:orthonormalexpansion}, may be chosen appropriately depending upon the desired accuracy of the numerical solution. 
The function \( \psi_n(x) \) is defined over a closed interval $\lsb{x_l, x_r}$. The Hamiltonian operator $H_w$ can be expressed as a $N \times N$ matrix (which we also denote as $H_w$) with  
the matrix element is given by 
\begin{equation} \label{eq:defHw}
 \lob{{H}_w }_{\alpha \beta}  =  \int_{x_l}^{x_r} \psi^{\star}_{\alpha} (x) \,  \hat{H}_w  \,  \psi_{\beta} (x) \, dx.
\end{equation}
The solution to \meqref{eq:defschrodoneDw} is given as 
\begin{equation} \label{eq:defpsifullexpansion}
    \Psi(x, t) = \sum_{n=1}^{N} d_n \exp\left(\frac{\lambda_n }{h_w} (t - t_f) \right) v_n(x),
\end{equation}
where $v_n(x) = \sum_{j=1}^{N}\, {{v}}_{n,j} \, \psi_j(x)  $
and ${\bf{v}}_n = [v_{n,1} \quad  v_{n,2} \quad \cdots \quad v_{n,N}]^T$ is an eigenvector of matrix $H_w$ with eigenvalue $\lambda_n$, and the constant $d_n$ is determined using the terminal condition given in \meqref{eq:terminal} as follows. 
First, using the orthonormality of  basis functions \( \psi_n(x) \)  and the terminal condition given in \meqref{eq:terminal}, we obtain:
\begin{equation}
    c_{\alpha}(t_f) = \int_{x_l}^{x_r}  \, \psi^{\star}_{\alpha} (x) \, \Psi(x, t_f) \, dx = \int_{x_l}^{x_r}  \, \psi^{\star}_{\alpha} (x) \,  R (x, t_f) \exp(-\frac{i}{h_r}  \Phi(x)) \, dx.
\end{equation}
Let ${\bf{c}}(t) = [c_1(t),\, c_2(t),\, \ldots,\, c_N(t)]$, ${\bf{d}} = [d_1,\, d_2,\, \ldots,\,d_N ]$ and 
$\mathcal{V} = [ {\bf{v}}_1 \, | \, {\bf{v}}_2 \, | \, \cdots \, | \, {\bf{v}}_N]$.
On computing $ \Psi(x, t_f)$ in two different ways using \meqref{eq:orthonormalexpansion} and \meqref{eq:defpsifullexpansion}, we obtain the following 
\begin{align}
\mathcal{V} \, {\bf{d}} = {\bf{c}}(t_f) \implies {\bf{d}} = \mathcal{V}^{-1} {\bf{c}}(t_f).
\end{align}

\subsubsection{An example problem} \label{sec:analyticexample}
Let us consider the following optimal control problem where the objective is to minimize the cost function
 \begin{align}
    J  = \frac{1}{2} C x^2(t_f) + \frac{1}{2} \int_{0}^{t_f} \, u^2 dt,
\end{align}
such that 
\begin{align}
\dot{x} = B u(t), 
\end{align}
and $x(0) = x_0$.
Here, $B$ and $C$ are positive real numbers, and the terminal time $t_f$ is given.  
In this case, 
\begin{align}
    \hat{H} = - \frac{h^2_r}{2} \lob{ B \,  \pdir{}{x}}^2,
\end{align}
and the Hamiltonian operator $\hat{H}_w$ is given by
\[
\hat{H}_w = -\hat{H} \Big|_{h_r = i {h}_w} = \frac{h^2_w}{2} \lob{ B \,  \pdir{}{x}}^2 
\]
This results in the modified \Schrodinger \, equation given by 
\begin{equation}
     \frac{\partial \Psi(x, t)}{\partial t} = \frac{h_w B^2}{2}   \, \lob{  \pddir{\Psi}{x}}.
\end{equation}
If we restrict to a finite domain $x \in [x_l,x_h]$,  a solution of the above equation is 
\begin{align}
 \Psi(x,t) =  \sum_{n=1}^{N} \, (C_{1n} e^{i\sqrt{\lambda_n}x} + C_{2n} e^{-i\sqrt{\lambda_n}x})\cdot e^{-  {h}_w \frac{B^2}{2} \lambda_n t},
\end{align}
where the specific values of $C_{1n}$, $C_{2n}$, and $\lambda_n$ could be determined by the initial and boundary conditions. 
For illustrating the nature of an approximate solution, we consider only one eigenvalue, say $\lambda_n$, and consider the wave equation of the following form
\begin{align}
    \Psi = C_{2n} \cdot e^{-  {h}_w \frac{B^2}{2} \lambda_n t} e^{-i\sqrt{\lambda_n}x}. 
\end{align}
On comparing the above with $\Psi = R(x,t)e^{i S(x,t) / h_w}$,
we obtain
\begin{align}
    i S(x,t) / {h}_w = -i\sqrt{\lambda_n} x \implies S(x,t) = - {{h}_w}\sqrt{\lambda_n} x.
\end{align}
Using \meqref{eq:uhat}, we have
\begin{align}
    u = B \pdir{S(x,t)}{x} =  - {B \sqrt{\lambda_n}}{{h}_w}.
\end{align}
The terminal condition (ref.~\meqref{eq:terminal}) implies that
\begin{align}
    S(x,t_f) = - \frac{1}{2} C x^2(t_f) \implies \sqrt{\lambda_n} \tilde{h}_w = \frac{1}{2} C x(t_f).  
\end{align}
Using the above, one can obtain 
\begin{align}
   u = - \frac{1}{2} B C x(t_f) \quad  x(t) = x_0 -  \frac{1}{2} B^2 C \, x(t_f) \, t, \quad \text{where } x(t_f) = \frac{2 x_0}{2 + B^2 C \, t_f}.  
\end{align}
 We note that the correct analytic solution in this case is 
\begin{align} \label{eq:analytic_sol_ex_one}
    u^{*} = - B C x(t_f) \quad 
    x^{*} =  x_0 -  B^2 C \, x(t_f) \, t, \quad \text{where } x(t_f) = \frac{ x_0}{1 + B^2 C \, t_f}.  
\end{align}
Here, we have used only one eigenvalue to obtain an approximate solution. The solution can be improved further by considering more eigenvalues as we show in 
\mref{sec:ex1} by considering 
a computational example with $B=C=t_f = x_0 = 1$.

\subsection{Multi-dimensional formulation}
\label{Sec:highdim}
We consider the following optimal control problem with the objective of minimizing the cost function
\begin{equation} \label{eq:costfunctionnewmultiD}
J = \Phi(\bm{x}(t_f)) + \int_{t_0}^{t_f} {L}(\bm{x}(t), \bm{u}(t)) \, dt ,
\end{equation}
subject to  the dynamical constraints
\begin{align}
     \dot{x_i} = f_i(x_1,\, \ldots ,\, x_n,\,  u_1, \ldots,\,   u_r), 
\end{align}
where $1 \leq i \leq n$ and 
$f_i(x_1,\, \ldots ,\, x_n,\,  u_1, \ldots   u_r)$ are functions of dynamic variables $x_j$ and control variables $u_j$.
Similar to \eqref{eq:extendenHT},
the extended Hamiltonian is given by
\begin{align}
    H_T = -L + f_i p_{x_i} + \alpha^i p_{u_i} + \beta^i {\theta}_{u_i},
\end{align}
where
$p_{u_i} = 0$ and ${\theta}_{u_i} = 0$ are primary and secondary constraints and
\begin{align}
    \theta = L - f_i \, p_{x_i}.
\end{align}
Further, 
${\theta}_{u_i} = \pdir{\theta}{u_i}$ and $\alpha^i$ and $\beta^i$ are arbitrary coefficient functions. 
Let 
\begin{equation}
    \phi_i = \begin{cases}
        & p_{u_i}  \quad \text{if } 1 \leq i \leq r, \\
        & \theta_{u_i} \quad \text{if } r+ 1 \leq i \leq 2r. 
    \end{cases}
\end{equation}
Let $M$ be a $2r \times 2r $ matrix such that the matrix element $M_{ij}$ is defined as 
\begin{align}
    M_{ij} = \pbr{\phi_i}{\phi_j},
\end{align}
where $ 1 \leq i, j \leq 2r$.
It can be verified that the matrix $M$ is given by
\begin{align}
  M =  \begin{bmatrix}
0_{r \times r} & -C^T \\
& \\
C & D
\end{bmatrix},
\end{align}
where $C$ and $D$ are $r \times r $ matrices with 
\begin{align}
    C_{ij} = \pbr{{\theta}_{u_i}}{p_{u_j}} =  {\theta}_{u_i u_j},
\end{align}
and 
\begin{align}
    D_{ij} = \pbr{{\theta}_{u_i}}{{\theta}_{u_j}} 
    =  \lob{{f_k}}_{u_i} {\theta}_{u_j x_k} - {\theta}_{u_i x_k} \lob{{f_k}}_{u_j},
\end{align}
for $i, j  \in \{1, \, \ldots, r\}$.
We note the following:
\begin{enumerate}[(i)]
\item $C$ is a symmetric matrix. Hence $C^{-1}$ is also a symmetric matrix.
\item $D$ is a skew-symmetric matrix.
\item $M$ is a skew-symmetric matrix. Hence $M$ can be expressed as
\begin{align}
     M =  \begin{bmatrix}
0_{r \times r} & -C \\
& \\
C & D
\end{bmatrix}.
\end{align}
\item 
As $M$ is skew-symmetric, $M^{-1}$ is also skew-symmetric.
\end{enumerate}
In this case, using the observation above, we have 
\begin{align}
    M^{-1} 
    = \bmat{ C^{-1} D {C}^{-1}} {C^{-1}} {-{C}^{-1}}{0_{r \times r}}.
\end{align}
For arbitrary functions \( g_1 \) and \( g_2 \), using the definition of the Dirac bracket and the structural properties of the matrices \( C \) and \( D \) as noted above, it can be shown that 
\begin{align}
\{ g_1, g_2 \}_{DB} &= \{ g_1, g_2 \} - \{ g_1, \phi_\alpha \} (M^{-1})_{\alpha\beta} \{ \phi_\beta, g_2 \} \nonumber \\
&= \{ g_1, g_2 \} - \{ g_1, p_{u_i}  \} (C^{-1} DC^{-1})_{ij} \{ p_{u_j}, g_2 \}
- \{g_1, p_{u_i} \} (C^{-1})_{ij} \{ \theta_{u_j}, g_2 \}
+ \{ g_1, \theta_{u_i} \} (C^{-1})_{ij} \{ p_{u_j}, g_2 \}. \label{eq:defNewDirac}
\end{align}
One can take \meqref{eq:defNewDirac} as a new definition of the Dirac bracket in the context of the optimal control problem under consideration. This new Dirac bracket depends upon the matrices $C$ and $D$ and provides a convenient and systematic method of computation of $\{ g_1, g_2 \}_{DB}$ for  arbitrary functions \( g_1 \) and \( g_2 \).
Using the above, we obtain the following Dirac bracket relations:
\begin{align}
    \dbr{x_\alpha}{{p_x}_j} &= \delta_{\alpha j} \\
 \dbr{x_\alpha}{{u}_j}   &= 
  (f_\alpha)_{u_k} \lob{C^{-1}}_{kj}\\
 \dbr{p_{x_\alpha}}{{u}_j} &= {\theta}_{x_\alpha u_k } 
 \lob{C^{-1}}_{kj} \\
  \dbr{u_\alpha}{u_j} &=  \lob{ C^{-1} D {C}^{-1}}_{\alpha j}.
\end{align}
Before proceeding further to obtain
the corresponding commutator relations, we recall that symmetrization operator for arbitrary  operators $\hat{g}_1$ and  $\hat{g}_2$ is defined as as 
$$\mwidetilde{\hat{g}_1 \hat{g}_2} = \frac{1}{2} \lob{\hat{g}_1 \hat{g}_2 + \hat{g}_2 \hat{g}_1}.
$$
Hence, we obtain 
the following commutator relations:
\begin{align}
    \cbr{\hat{x}_\alpha}{{\hat{p}_{x_j}}} &=  i h_r~  \delta_{\alpha j} \\
  \cbr{\hat{x}_\alpha}{\hat{u}_j}   &= i h_r~ 
  \mwidetilde{(f_\alpha)_{u_k} \lob{C^{-1}}_{kj} } \label{eq:xiuj}\\
 \cbr{\hat{p}_{x_\alpha}}{\hat{u}_j} &= i h_r~  \mwidetilde{{\theta}_{x_\alpha u_k } 
 \lob{C^{-1}}_{kj} } \label{eq:pxiuj}\\
  \cbr{\hat{u}_\alpha}{\hat{u}_j} &=  i h_r~  \mwidetilde{\lob{ C^{-1} D {C}^{-1}}_{\alpha j}}. \label{eq:uiuj}
\end{align}
From the preceding discussion, it immediately follows that
\begin{align}
    \hat{p}_{x_j} = \frac{h_r}{i} \pdir{}{x_j}.
\end{align}
Next, we express $\hat{u}_j$ as
\begin{align}
    \hat{u}_j = {\mathcal{M}}_{jk} \pdir{}{x_k} + {\mathcal{N}}_{j},
\end{align}
where ${\mathcal{M}}_{jk}$ and ${\mathcal{N}}_{j}$ are unknown functions to be determined. 
Then, on using \meqref{eq:xiuj} on a test function $g$ we obtain,
\begin{align}
    (\hat{x}_\alpha \hat{u}_j - \hat{u}_j \hat{x}_\alpha) g =  {x}_\alpha \lob{{\mathcal{M}}_{jk} \pdir{}{x_k} + {\mathcal{N}}_{j}} g - \lob{{\mathcal{M}}_{jk} \pdir{}{x_k} + {\mathcal{N}}_{j}} (x_\alpha g) = - {\mathcal{M}}_{j \alpha} \, g =   i h_r~ 
  \mwidetilde{(f_\alpha)_{u_k} \lob{C^{-1}}_{kj} } g 
\end{align}
It is clear that
\begin{align}
    {\mathcal{M}}_{j\alpha} = -i h_r~ 
  \mwidetilde{(f_\alpha)_{u_k} \lob{C^{-1}}_{kj} }, 
\end{align}
or equivalently
\begin{align}
     {\mathcal{M}}_{jk} = -i h_r~ 
  \mwidetilde{(f_k)_{u_\alpha} \lob{C^{-1}}_{\alpha j} }. 
\end{align}
Similarly, we obtain the following form \meqref{eq:pxiuj}: 
\begin{align*}
 \lob{\hat{p}_{x_\alpha} \hat{u}_j - \hat{u}_j  \hat{p}_{x_\alpha}} g & =  \frac{h_r}{i} \pdir{}{{x}_\alpha} \lob{{\mathcal{M}}_{jk} \pdir{}{x_k} + {\mathcal{N}}_{j}} g - \lob{{\mathcal{M}}_{jk} \pdir{}{x_k} + {\mathcal{N}}_{j}} \lob{\frac{h_r}{i} \pdir{g}{{x}_\alpha}} \\
 &= \frac{h_r}{i}  \lob{ \pdir{{\mathcal{M}}_{jk}}{x_\alpha} \pdir{g}{x_k} + \pdir{{\mathcal{N}}_{j}}{x_\alpha} g}  = i h_r ~  \mwidetilde{{\theta}_{x_\alpha u_k } 
 \lob{C^{-1}}_{kj}} g.
\end{align*}
This implies, 
\begin{align}
  \pdir{{\mathcal{N}}_{j}} {x_\alpha} = - \lob{ \mwidetilde{{\theta}_{x_\alpha u_k } 
 \lob{C^{-1}}_{kj} }  + \pdir{{\mathcal{M}}_{jk}}{x_\alpha} \pdir{}{x_k}}.   
\end{align}
Next, we restrict to the following 
case.
Let the Lagrangian be given by
\begin{equation}
    \label{eq:lagrangian u-R-Vcost formulation}
    L= \frac{1}{2} m_{k j} \,  u_{k} u_j + V(x_1,\, \ldots ,\, x_n),
\end{equation}
where $ 1 \leq k ,j \leq r$, and  $x_1$, $x_2$, $\ldots$, $x_n$ are the state variables, and $u_k $ are control variables such that 
\begin{equation}
    \label{eq:xdot-g-F representation}
    \begin{split}
        \dot{x}_\alpha  = f_\alpha (x_1,\, \ldots ,\, x_n,\,  u_1, \ldots,\,   u_r) = a_\alpha (x_1,\, \ldots ,\, x_n) + b_{\alpha  k} (x_1,\, \ldots ,\, x_n,)  u_{k}.
    \end{split}
\end{equation}
We also assume that $m_{ij} = m_{ji}$.
The Hamiltonian $H$ is given by
\begin{align} \label{eq:multidimprobbH}
    H =   -L + f_\alpha  p_{x_\alpha } =  - \, \frac{1}{2} m_{\alpha j} \,  u_\alpha  u_j -  V(x_1,\, \ldots ,\, x_n) +  a_\alpha (x_1,\, \ldots ,\, x_n) p_{x_\alpha } + b_{\alpha  k} (x_1,\, \ldots ,\, x_n)  u_{k} p_{x_\alpha }.
\end{align}
It is easy to check that
\begin{align}
    \theta &= L - f_\alpha  p_{x_\alpha } = \frac{1}{2} m_{k j} \,  u_{k} u_j + V(x_1,\, \ldots ,\, x_n)   - a_\alpha (x_1,\, \ldots ,\, x_n) p_{x_\alpha } - b_{\alpha  k} (x_1,\, \ldots ,\, x_n)  u_{k} p_{x_\alpha }, \\
    \theta_{u_k} &= m_{\alpha k} u_\alpha - b_{\alpha k} \, p_{x_\alpha}, \\
    C_{jk} &= \theta_{u_k u_j} = m_{jk}.
\end{align}
Next, we compute
\begin{align}
    {\mathcal{M}}_{jk} =  -i h_r~ 
  \mwidetilde{(f_k)_{u_\alpha } \lob{C^{-1}}_{\alpha j}}  =  -i h_r~ \mwidetilde{b_{k\alpha } \lob{C^{-1}}_{\alpha j}}, 
\end{align}
and
\begin{align}
    \theta_{x_\alpha  u_k} = -  (b_{\beta k})_{x_\alpha } p_{x_\beta}. 
\end{align}
Therefore,
\begin{align*}
  \pdir{{\mathcal{N}}_{j}} {x_\alpha } &= - \lob{ \mwidetilde{{\theta}_{x_\alpha  u_k } 
 \lob{C^{-1}}_{kj} }  + \pdir{{\mathcal{M}}_{jk}}{x_\alpha } \pdir{}{x_k}} \\
 &=        \frac{ \lob{C^{-1}}_{kj} \lob{
 (b_{\beta k})_{x_\alpha } p_{x_\beta} +   p_{x_\beta} (b_{\beta k})_{x_\alpha } } }{2}  
  -   \lob{C^{-1}}_{\beta j} {(b_{k\beta})}_{x_\alpha } p_{x_k}    \\
  & =    \frac{ \lob{C^{-1}}_{kj} \lob{
 (b_{\beta k})_{x_\alpha } p_{x_\beta} +   p_{x_\beta} (b_{\beta k})_{x_\alpha } } }{2}  
  -   \lob{C^{-1}}_{k j} {(b_{\beta k })}_{x_\alpha } p_{x_\beta}  \\ 
  & =  \frac{ \lob{C^{-1}}_{kj} \lob{ -
 (b_{\beta k})_{x_\alpha } p_{x_\beta} +   p_{x_\beta} (b_{\beta k})_{x_\alpha } } }{2}  
  \\ 
 & =  - i h_r~   \frac{\lob{C^{-1}}_{kj} (b_{ \beta k})_{x_\beta x_\alpha }}{2}.   
\end{align*}
It follows that
\begin{align}
    N_j =  - i h_r~   \frac{\lob{C^{-1}}_{kj} (b_{ \beta k})_{x_\beta }}{2}. 
\end{align}
This gives
\begin{align}
    \hat{u}_j = - i h_r
    \lob{  b_{k \alpha} \lob{C^{-1}}_{\alpha j}  \pdir{}{x_k} + \lob{C^{-1}}_{kj} \frac{(b_{\beta k})_{x_\beta}}{2} } = - i h_r \lob{C^{-1}}_{k j}
    \lob{  b_{\beta k}   \pdir{}{x_\beta} +  \frac{(b_{\beta k})_{x_\beta}}{2} }.
\end{align}
Using the expression obtained above for $\hat{u}_j$ and $\hat{p}_{x_j}$
in \meqref{eq:multidimprobbH}, the expression for the Hamiltonian operator $\hat{H}$ can easily be obtained. The corresponding \Schrodinger \ equation can easily be set up, and its solution can be obtained using a simple extension of the spectral method described in \mref{Sec:spectraloneD} to higher dimensions.
Suitable orthonormal basis functions such as sine, cosine, Chebyshev, Legendre or Jacobi basis functions that satisfy the boundary conditions in higher dimensions can be used for spectral approximation. 

\section{Variational quantum algorithms (VQAs)}
\label{sec:VQA}
Variational quantum algorithms (VQAs) provide a promising approach to leverage near-term quantum devices for applications such as quantum chemistry, quantum simulation, and machine learning \cite{Cerezo_2021}. VQAs consist of two main components: a parametrized quantum circuit that prepares a trial state $ \lvert \psi (\theta) \rangle$, and a classical optimizer that updates the parameters $\theta$ to minimize a cost function $C(\theta)$ that depends on the expectation value of some observable $\langle \psi(\theta)|H|\psi(\theta)\rangle$, where $H$ is the Hamiltonian of the system of interest. The goal of a VQA is to find an optimal state that approximates the ground state of $H$ or to solve some other optimization problem related to $H$.

An ``ansatz'' refers to the parameterized form of the quantum circuit that serves as a guess or approximation for the solution to a given quantum problem and is used to represent and manipulate quantum states during computation. 
The quantum circuit can be designed using various ansatzes, such as hardware-efficient ansatzes, problem-specific ansatzes, or machine learning-inspired ansatzes \cite{Qin_2023}. The classical optimizer in VQAs can be chosen from a variety of optimization algorithms, such as gradient-based methods, derivative-free methods, or meta-heuristics \cite{wurtz2021classically, pellow2021comparison}. The choice of the quantum circuit and the classical optimizer depends on the problem at hand, the available quantum resources, and the noise characteristics of the quantum device.

Some examples of VQAs are the variational quantum eigensolver (VQE) \cite{Peruzzo_2014}, which aims to find the ground state energy of a Hamiltonian; the quantum approximate optimization algorithm (QAOA) \cite{farhi2014quantum}, which seeks the optimal solution of a combinatorial optimization problem; and the quantum neural network (QNN) \cite{Mari_2020}, which performs machine learning tasks using quantum circuits as trainable models. VQAs have been demonstrated on various quantum platforms and have shown potential for achieving quantum advantage in the near future \cite{Arute_2019}.

\subsection{VQE for non-Hermitian systems}
\label{subsec:VQA_non_Hermitian}
Non-Hermitian systems have gained significant attention in recent years due to their unique topological properties, which have no Hermitian counterparts \cite{PhysRevX2018, RevModPhys2021}. These systems exhibit exceptional degeneracies, where both eigenvalues and eigenvectors coalesce, leading to phenomena drastically distinct from the familiar Hermitian realm \cite{RevModPhys2021}. The role of topology in non-Hermitian systems has far-reaching physical consequences observable in a range of dissipative settings \cite{RevModPhys2021}. A coherent framework for understanding the topological phases of non-Hermitian systems has been developed, providing insights into the role of topology in these systems \cite{PhysRevX2018}. 

The conventional Variational Quantum Eigensolver (VQE) algorithm is designed to solve eigenstates of many-body systems and minimize the variational energy or expectation value \cite{Tilly_2022}. However, for non-Hermitian systems, the eigenvalues may not be real numbers, rendering the conventional VQE inapplicable. Therefore, new approaches are required to address non-Hermitian systems. For the computation of eigenvalues in our non-Hermitian systems, we have utilized a modified version of the approach presented in \cite{Xie2024}. We note that the approach in \cite{Xie2024} is tailored for non-Hermitian systems, and it employs the principle of variance minimization. 
The method proposed in \cite{Xie2024} involves initially constructing Hermitian Hamiltonians $\hat{A}$ and $\hat{A'}$ from a given non-Hermitian Hamiltonian $\hat{H}$. Here, $\hat{H}$ and $\hat{H}^{\dag}$ represent distinct eigenstates, referred to as right and left eigenstates, respectively. Similarly, $\hat{A}$ and $\hat{A'}$ are employed for solving right and left eigenstates, respectively. They are defined as follows:

\begin{equation}
\begin{split}
    \label{eq: hermitian_formulation_eqn_vqa}
    \hat{A}(E) &= (\hat{H}^{\dag} - \bar{E})(\hat{H}-E), \\
    \hat{A'}(E) &= (\hat{H}-E)(\hat{H}^{\dag} - \bar{E}),
\end{split}
\end{equation}
where $\bar{E}$ represents the complex conjugate of an eigenvalue $E$.
The algorithm relies on constructing a cost function that minimizes the expected value of $\hat{A}$ with respect to variable $E$ and a given state $ \lvert \psi (\theta) \rangle$. Here, $ \lvert \psi (\theta) \rangle = U( \theta) \ket{0}$, where a Parameterized Quantum Circuit (PQC) constructs the unitary operator $U(\theta)$. We note that $\hat{A}(E)$ equals zero if and only if $\lvert \psi (\theta) \rangle$ is the right eigenstate and $E$ is an eigenvalue of the Hamiltonian $\hat{H}$. Therefore, the cost function $\mathcal{C}(\theta, E_r, E_i)$ is formulated to find the expected value of $\hat{A}(E)$, with $\theta$ representing the PQC parameters and $E=E_r+iE_i$ as a parameter, where $E_r$ and $E_i$ denote the real and imaginary components of $E$, respectively. 

The above-described algorithm (given in \cite{Xie2024}) was used with slight modifications for the computational examples described in Example 6 in \mref{sec:Simulations}. The modifications include the use of a different ansatz that is similar to the Efficient-SU2 ansatz proposed by authors in \cite{Kandala_2017}, with fewer entangling layers.

\section{Computational examples}
\label{sec:Simulations}
In this section, we will illustrate the application of our proposed framework using two computational examples assuming access to a perfect VQE for non-Hermitian systems. 
We consider two types of orthonormal basis functions in our computational examples, namely the sinusodial basis functions and  Legendre basis functions. Other orthonormal basis functions including Jacobi and Chebyshev polynomials may also be used.
The dependence of numerical solutions on various parameters, including the number of basis functions, $h_w$, and the nature of eigenvalues of the Hamiltonian are explored. Numerical solutions are compared with exact solutions (where available).

First, we will consider computational examples of the following form, which is 
a special case of the problem given in \meqref{eq:defL} and \meqref{eq:defxdot}.
The objective is to find a control function \( {u}^*(t) \) that minimizes the given performance index \( J \), 
\begin{equation} \label{eq:Lcostfunction}
J = \frac{1}{2} C \, (x(t_f))^2 + \frac{1}{2} \int_{t_0}^{t_f} (u(t))^2 dx,
\end{equation}
 subject to the dynamics constraint
 \begin{equation}
    \label{eq:Lc_paper}
\dot{x}{(t)}  =  A x(t)  + 
B u(t), \quad t \in [t_0,t_f],
\end{equation}
with the initial condition  ${x}(t_0) = x_0$. 

Assume $A$, $B$, and $C>0$ are real constants, with $A$ and $B$ not equal to zero simultaneously.
In this case, one can easily obtain the optimal solution analytically. The expressions for the state and the control variables for the optimal solution are as follows \cite{murray2009optimization}:
\begin{align}
    x^*(t) &= e^{A(t - t_0)} \,  x_0 + \frac{B^2 C \,  e^{A(t_f - t_0)} x_0 }{2A - B^2 C \left(1 - e^{2A(t_f - t_0)}\right)} \left( e^{A(t_f - t)} - e^{A(t + t_f - 2t_0)} \right), \label{eq:xstar} \\
     u^*(t) &= -\frac{2ABC \,  e^{A(2t_f - t - t_0)} \, x_0}{2A - B^2\, C \left(1 - e^{2A(t_f - t_0)}\right)}. \label{eq:ustar} 
\end{align}
For $A=0$ case, the optimal solution can be obtained directly or by taking $\lim A \to 0 $ in \meqref{eq:xstar} and \meqref{eq:ustar}, and it results in the optimal solution given in \meqref{eq:analytic_sol_ex_one} (with $t_0 =0$).

We will use our framework described in \mref{Sec:Canonical_quantization} to solve this optimal control problem.
We will consider the domain
$ [x_l,x_r] = [-L_x,L_x]$ and
use the following $N = N_c + N_s$ basis functions defined as 
\begin{align}
\psi_{c,k}(x) &=    \cosbasisoned{x},  \quad \text{for } k = \iseqp{N_c}, \\
\psi_{s,k}(x)  &=   \sinbasisoned{x}, \quad \text{for } k = \iseqp{N_s}, 
\end{align}
in Examples $1-4$  and Example $6$. 
We will use both sinusoidal and Legendre polynomial basis functions in Example $3$.
In Example $5$, the following basis functions will be used:
\begin{align}
    \chi_{ss,k_1,k_2} (x_1, x_2) =   \psi_{s,k_1}(x_1) \psi_{s,k_2}(x_2), \quad k_1 = \iseqp{N_{1,s}}, \quad k_2 = \iseqp{N_{2,s}} \\
    \chi_{sc,k_1,k_2} (x_1, x_2) =   \psi_{s,k_1}(x_1) \psi_{c,k_2}(x_2), \quad k_1 = \iseqp{N_{1,s}}, \quad k_2 = \iseqp{N_{2,c}}\\
    \chi_{cs,k_1,k_2} (x_1, x_2) =   \psi_{c,k_1}(x_1) \psi_{s,k_2}(x_2), \quad k_1 = \iseqp{N_{1,c}}, \quad k_2 = \iseqp{N_{2,s}} \\
    \chi_{cc,k_1,k_2} (x_1, x_2) =   \psi_{c,k_1}(x_1) \psi_{c,k_2}(x_2), \quad k_1 = \iseqp{N_{1,c}}, \quad k_2 =\iseqp{N_{2,c}},
\end{align}
where 
\begin{align}
\psi_{s,k_j}(x_j)  &=   \sinbasis{j}, \quad \text{for } k_j = \iseqp{N_{j,s}}, \\
\psi_{c,k_j}(x_j) &=    \cosbasis{j},  \quad \text{for } k_j = \iseqp{N_{j,c}},
\end{align}
 on the domain $-L_j\leq x_j \leq L_j$ for $j=1, \, 2$.

\subsection{Example 1}  \label{sec:ex1}
\begin{figure}[tbh]
    \centering
    \includegraphics[scale=0.75]{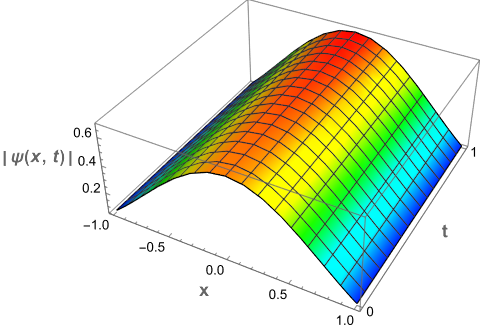}
    \caption{Wave function for Example 1, with $N_c=N_s=2$ (i.e., $N=4$) and $h_w = 0.1$.}
    \label{fig:wave_fn_ex1}
\end{figure}

For our first example, we consider a special case of the problem described in \meqref{eq:Lcostfunction} and \meqref{eq:Lc_paper} with the following parameters.
\begin{align}
    \label{eq:1dexone}
A = 0, \quad
B = 1, \quad
C = 1, \quad
t_0 = 0, \quad
t_f = 1, \quad 
x_0 = 1.
\end{align}
This example also corresponds to the example problem discussed in \mref{sec:analyticexample} with $B=C=t_f = x_0 = 1$.
In this case, we fix the following parameters
\begin{align}
L_x = 2,
 \quad
h_w = 0.1, \quad R(x, t_f) = 1.
\end{align}
Further, we obtain optimal solutions for the state and control variables using our proposed approach for $N_s = N_c= 2$, as described below.
In this case, we obtain the Hamiltonian matrix ${H}_w$ as the diagonal matrix
${H}_w = \operatorname{diag}(0.00308425,\ 0.0277583,\ 0.012337,\ 0.049348)$  (ref.~\eqref{eq:defHw}).
Further, the following wave function was obtained.
$$\Psi(x,t) = 0.384295 \, e^{0.0308425 (t-1)} \cos (\frac{\pi}{4}x)+0.300267 \, e^{0.277583 (t-1)} \cos (\frac{3\pi }{4} x).$$
Next, using the approach outlined in Section~\ref{Sec:oneDProb} (see Eq.~\ref{eq:defufinal}), we computed the optimal state and control solutions, and obtained the PI value of $0.275925$.  
When the same method is applied with \(N_s = N_c = 4\), the resulting PI value is $0.25098$, which closely matches the analytic optimum of $0.25$.  
Note that the accuracy of the optimal solution obtained using our method depends on the chosen parameters \(x_l\), \(x_r\), \(L_x\), \(h_w\), \(L_x\), \(N_s\), and \(N_c\). For instance, if we consider $[x_l,x_r] = [-1,1]$ for this problem, keeping all the other parameters the same, then we obtain the following Hamiltonian matrix. 
\begin{equation}
{H}_w = 
\begin{bmatrix}
 0.002523870 & 0.00883573 & 0 & 0 \\
 0.000981748 & 0.01093390 & 0 & 0 \\
 0 & 0 & 0.00616850 & 0.020944 \\
 0 & 0 & 0.00523599 & 0.024674
\end{bmatrix}.
\end{equation}
We note that, in this case, the basis functions are 
\[
\frac{1}{\sqrt{2}} \left\{
\cos\left( \frac{\pi}{4} x \right),\ 
\cos\left( \frac{3\pi}{4} x \right),\ 
\sin\left( \frac{\pi}{2} x \right),\ 
\sin\left( \pi x \right)
\right\}
\]
and they are not orthonormal on the interval \([-1,1]\). However, as seen above, the off-diagonal terms in the matrix \({H}_w\) are small. Therefore, we proceed as follows to obtain an approximate solution. 
The matrix ${H}_w$ has the following eignevalues
\begin{align}
    \lambda_1 = 0.0293954, \quad \lambda_2 = 0.0118627, \quad \lambda_3 = 0.00159502, \quad \lambda_4 =  0.00144715,
\end{align}
such that the matrix $\mathcal{V}$ contains the corresponding eigenvectors as columns
\begin{equation}
\mathcal{V} = 
\begin{bmatrix}
\mid & \mid & \mid & \mid \\
{\bf{v}}_1 \, & \, {\bf{v}}_2 \, & \,{\bf{v}}_3 \, & \, {\bf{v}}_4  \\
\mid & \mid & \mid & \mid \\
\end{bmatrix}
=
\begin{bmatrix}
 0 & 0 & 0.669667 & 0.742661 \\
 0.687268 & 0.726404 & 0 & 0 \\
 0.99452 & -0.104549 & 0 & 0 \\
 0 & 0 & 0.97552 & -0.21991
\end{bmatrix}.
\end{equation}
Next, we obtain the following
\begin{align}
    [c_1(1),\, c_2(1),\, c_3(1),\, c_4(1)] \approx [0.5429,\, 0.42536,\, 0,\, 0 ], \quad 
    [d_1,\, d_2,\, d_3,\, d_4] \approx [0,\, 0.604058,\, 0.128454,\, 0 ].
\end{align}
Then, the wave function is given by
\begin{equation} \label{eq:defpsifullexpansionex}
    \Psi(x, t) \approx \sum_{n=1}^{4} d_n \exp\left({10 \lambda_n } (t - 1) \right) v_n(x).
\end{equation}
A plot of $ \lvert \Psi(x, t)\rvert$ is
shown in \mfig{fig:wave_fn_ex1}.

\begin{figure}[H]
\centering
\begin{subfigure}{.5\textwidth}
  \centering
  \includegraphics[width=.85\linewidth]{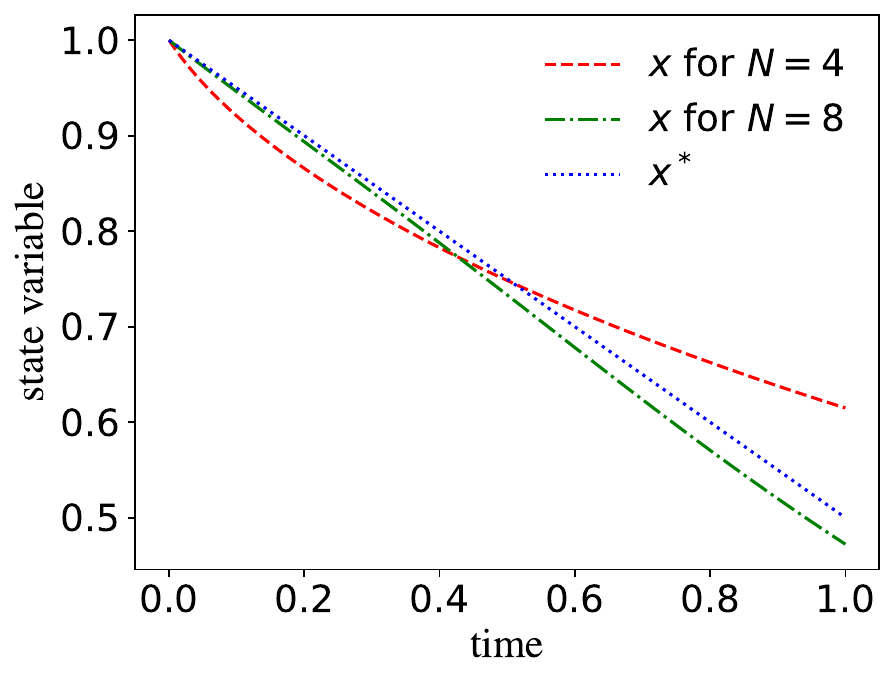}
  \caption{Time evolution of state variable}
\end{subfigure}%
\begin{subfigure}{.5\textwidth}
  \centering
  \includegraphics[width=.85\linewidth]{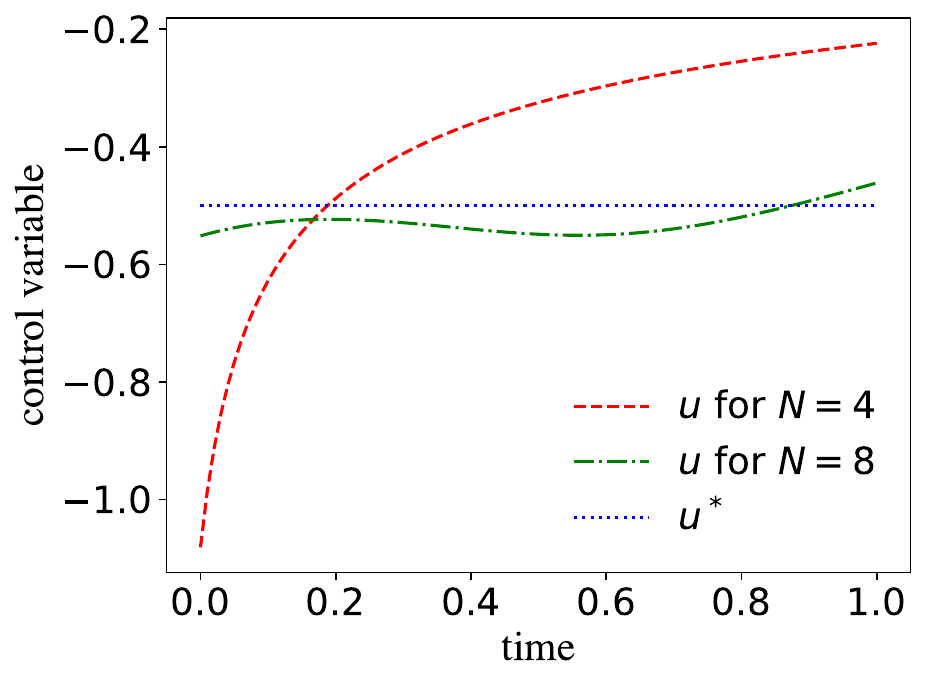}
  \caption{Time evolution of control variable}
\end{subfigure}
\caption{Time evolution of the state and control variables for Example 1 (with $h_w=0.1$). Results from the proposed framework for $N=4$ case (dashed lines) and $N=8$ case (dashed-dotted lines) are shown, including a comparison with analytical solutions (shown in dotted lines).}
\label{fig:ex_analytic}
\end{figure}

Next, using our approach described in Section~\ref{Sec:oneDProb} (see Eq.~\ref{eq:defufinal}), we obtained the optimal solutions for the state and control variables, as illustrated in \mfig{fig:ex_analytic}. In this case, the PI value was found to be 0.277732.
Applying the same approach for \(N_s = N_c = 4\) yields a PI value of 0.250998, which is in good agreement with the analytic optimal PI value of 0.25. It is important to note that the quality of the optimal solution obtained using our proposed approach depends on the chosen values of \(h_w\), \(L_x\), \(N_s\), and \(N_c\). The time evolution of the state and control variables, along with the correct analytic solutions (ref.~\meqref{eq:analytic_sol_ex_one}), are shown in \mfig{fig:ex_analytic}. It is evident from the plots that the results obtained via our proposed approach are in good agreement with the corresponding analytic solutions.

\subsection{Example 2}
\label{Sec:example2}

\begin{figure}[tbh]
    \centering
    \includegraphics[scale=0.75]{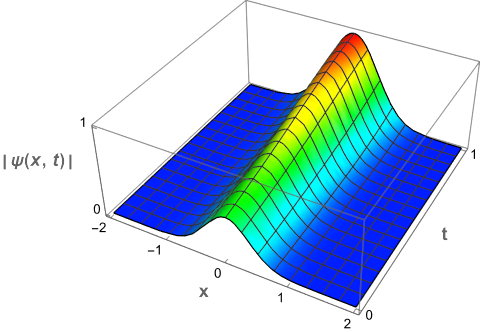}
    \caption{Wave function amplitude for Example 2,  with $N_s=N_c=8$ (i.e., $N=16$) and $h_w = 2$.}
    \label{fig:wave_fn_ex2}
\end{figure}

For our second example, we consider the special case of ~\meqref{eq:Lcostfunction} and \meqref{eq:Lc_paper} with 
\begin{align}
    \label{eq:1dextwo}
A = 2, \quad
B = \frac{1}{2}, \quad
C = 10, \quad
t_0 = 0, \quad
t_f = 1, \quad 
x_0 = 1.
\end{align}
In this case, we fix the following
\begin{align}
L_x =2,
 \quad
N_s = N_c= 8 \, \, (\text{i.e.},\, N = 16), \quad
h_w = 2, \quad R(x, t_f) = 1.
\end{align}
With the above setup, on using our approach described in \mref{Sec:Canonical_quantization} (also illustrated in more detail in the previous example, i.e., Example 1), we obtain a PI of $7.913042$, whereas the analytic PI is $7.913041$. 
The wave-function plot is shown in Figure~\ref{fig:wave_fn_ex2}. 

\begin{figure}[tbh]
\centering
\begin{subfigure}{.5\textwidth}
  \centering
  \includegraphics[width=.85\linewidth]{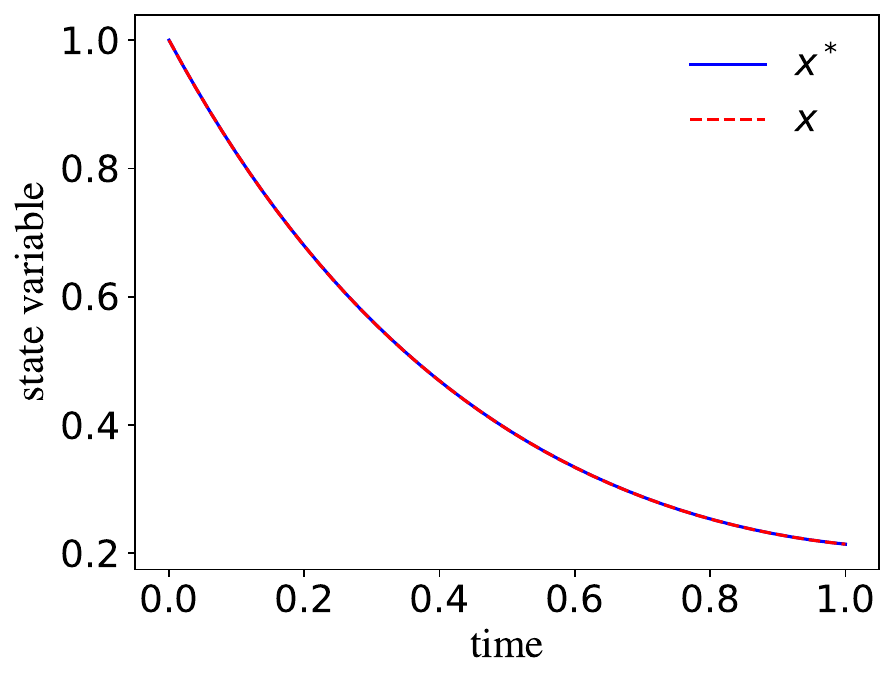}
  \caption{Time evolution of state variable}
  \label{fig:state_a_2}
\end{subfigure}%
\begin{subfigure}{.5\textwidth}
  \centering
  \includegraphics[width=.85\linewidth]{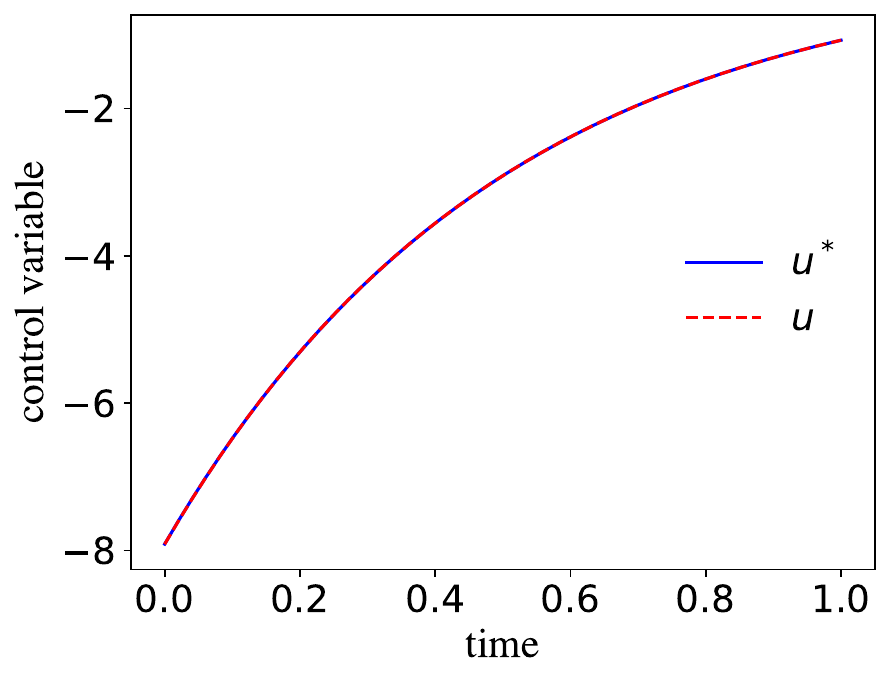}
  \caption{Time evolution of control variable}
  \label{fig:control_a_2}
\end{subfigure}
\caption{Time evolution of state and control variables for Example 2, with $N_s=N_c=8$ (i.e., $N=16$) and $h_w = 2$. Results from the proposed framework (shown in dotted lines) are compared with analytical solutions (shown in solid lines).}
\end{figure}

The plots showing the time evolution of the state and control variables, along with the corresponding correct analytic solutions (ref.~\meqref{eq:xstar} and \meqref{eq:ustar}), are shown in Figure~\ref{fig:state_a_2} and Figure~\ref{fig:control_a_2} respectively. The plots clearly show that the results from our proposed method closely match the corresponding analytical solutions.

\subsection{Example 3}
\label{subsubsec: 1d_example_with_gibbs}
For our third example, we consider the special case of ~\meqref{eq:Lcostfunction} and \meqref{eq:Lc_paper} with 
\begin{align}
    \label{eq:1dexthree}
A = -2, \quad
B = \frac{1}{2}, \quad
C = 10, \quad
t_0 = 0, \quad
t_f = 1, \quad 
x_0 = 1.
\end{align}
In this case, we fix the variables $L_x = 2$ and $R(x, t_f)=1$ and investigate the effects of varying the parameters $h_w$ and $N$, along with the type of basis functions. 

\begin{figure}[tbh]
    \centering
    \includegraphics[scale=0.75]{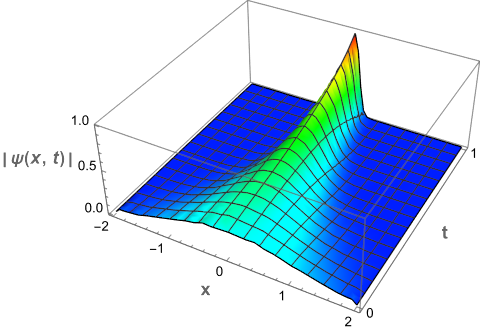}
    \caption{Wave function amplitude for Example 3,  with $N_s=N_c=16$ (i.e., $N=32$) and $h_w = 0.1$.}
    \label{fig:wave_fn_ex3}
\end{figure}

The plot of wave-function amplitude (for with $N_s=N_c=16$ (i.e., $N=32$) and $h_w = 0.1$) is shown in Figure~\ref{fig:wave_fn_ex3}.
Table \ref{tab:c_1d_a_-2} shows the dependence of the performance index on $h_w$ and $N$ using sinusoidal basis functions. 
From the theoretical considerations discussed in \mref{Sec:Canonical_quantization}, we expect that the result of our proposed approach matches with the corresponding correct optimal solution in the limit $h_w \rightarrow 0$ and $N \rightarrow \infty$. The results shown in Table 1 are consistent with this expectation. However, we note that for fixed value of $h_w$, the error in the performance index initially decreases with increasing $N$ but subsequently increases with $N$. In other words, for any fixed $h_w$ there exists an optimum value of $N$ that gives the best solution (i.e. with the least error in performance index) that cannot be further improved by increasing $N$ (ref.~ $N=16$ and $N=32$ cases for $h_w=1$ in Table \ref{tab:c_1d_a_-2}). Instead, better solutions can be obtained by decreasing $h_w$ and then increasing $N$ until the errors are no longer decreasing.

\begin{figure}[H]
\centering
\begin{subfigure}{.5\textwidth}
  \centering
  \includegraphics[width=.85\linewidth]{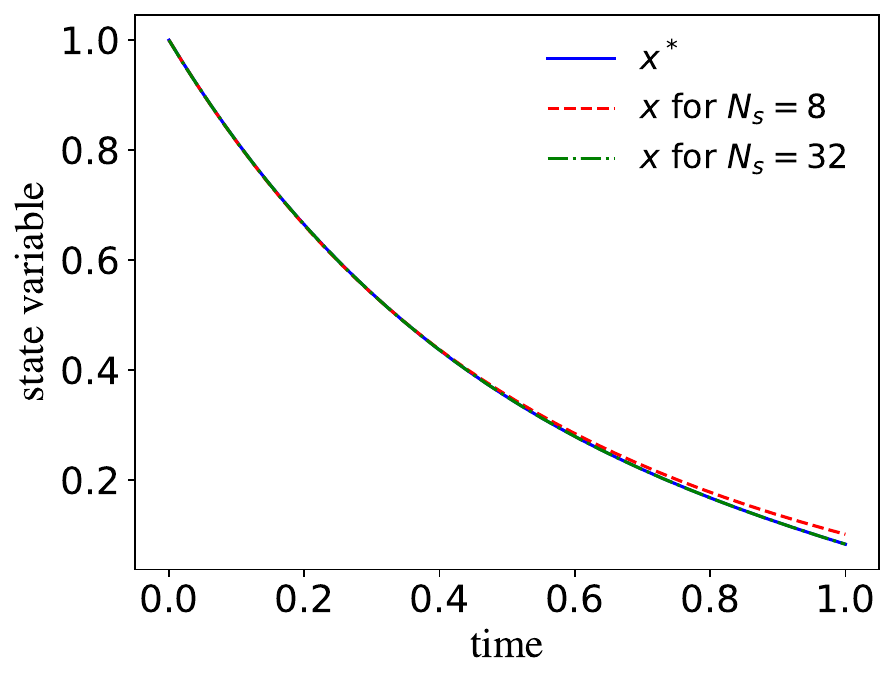}
  \caption{Time evolution of state variable}
  \label{fig:c_gibbs_state}
\end{subfigure}%
\begin{subfigure}{.5\textwidth}
  \centering
  \includegraphics[width=.85\linewidth]{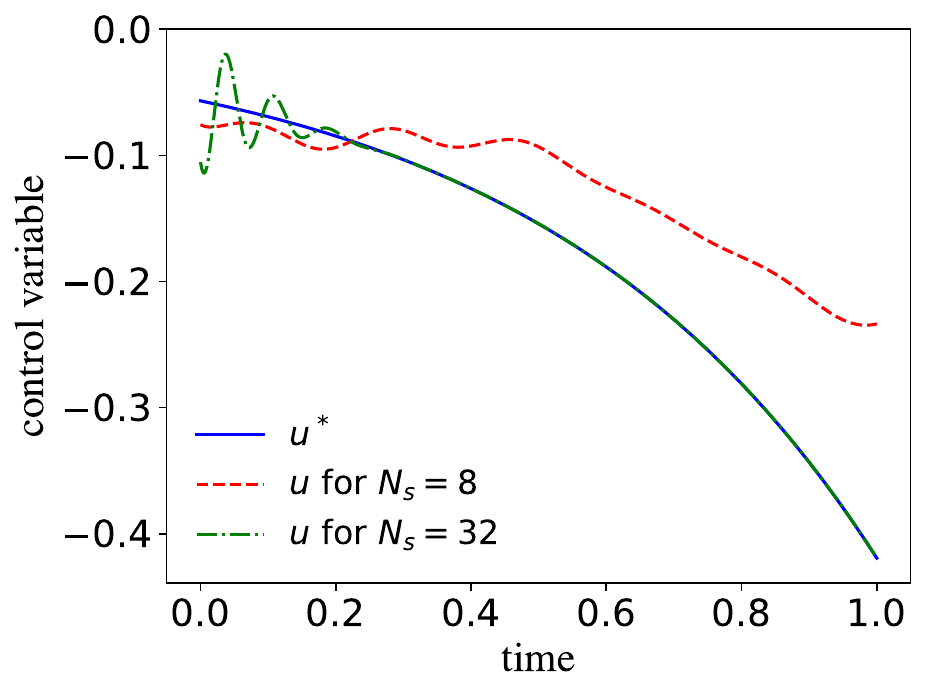}
  \caption{Time evolution of control variable}
  \label{fig:c_gibbs_control}
\end{subfigure}
\caption{Time evolution of the state and control variables for Example 3 in \mref{sec:Simulations} (with $h_w=0.1$). Results from the proposed framework for $N_s=8$ (or $N=16$)  and $N_s=32$ (or $N=64$) cases are shown in dashed and dashed-dotted lines respectively. Corresponding analytical solutions are depicted in solid lines.}
\label{fig:gibbs}
\end{figure}

 Plots showing the time evolution of the state and control variables obtained from our proposed method (for $h_w = 0.1$ and $N_s = 8,\, 32$, using sinusoidal basis functions), along with the corresponding correct analytic solutions (ref.~\meqref{eq:xstar} and \meqref{eq:ustar}), are shown in \mfig{fig:c_gibbs_state} and  
 \mfig{fig:c_gibbs_control}
 respectively. The plot for the time evolution of the control variable shows oscillatory behavior (akin to Gibbs phenomenon). The percentage errors in performance index for these cases can be noted from Table \ref{tab:c_1d_a_-2}. We note that the percentage error in the performance index is given by  $ \frac{100 (PI - PI^*) }{PI^*} $, where $PI^*$ is the correct analytic optimal performance index and $PI$ is the performance index obtained using our proposed approach.

\begin{table}[H]
\centering
\caption{Comparison of percentage errors in Performance Index (PI) for different $h_w$ and different number ($N=2 N_s$) of sinusoidal basis functions. Here, we consider the cases for $N_s = N_c$, i.e $N = N_s + N_c = 2 N_s$, with $L_x =2$.}
\label{tab:c_1d_a_-2}
\begin{tabular}{lcl}
\toprule
$h_w$ & $N_s (=N_c)$ & PI error (\%) \\
\midrule
\midrule
0.5  & 4 & $ 4.5 \times 10^1$ \\ 
0.5  & 8 & $ 2.9 \times 10^1$ \\ 
0.5 & 16  &  $1.9 \times 10^1$\\ 
0.5 & 32  &  $4.5 \times 10^0$\\ 
0.1  & 4  & $3.67 \times 10^{1}$ \\
0.1  & 8   & $7.91 \times 10^{0}$ \\ 
0.1  & 16 & $7.24 \times 10^{-2}$\\ 
0.1  & 32 & $8.10 \times 10^{-2}$ \\ 
0.05 & 4   & $4.69 \times 10^{1}$ \\
0.05 & 8   & $2.09 \times 10^{1}$ \\
0.05 & 16   & $3.50 \times 10^{-1}$ \\
0.05 & 32 & $4.58 \times 10^{-4}$ \\ 
\bottomrule
\end{tabular}%
\end{table}

From the plot in \mfig{fig:wave_fn_ex3}, we see that the wave function is mostly concentrated in the region $x \in [-1,1]$. We explored the solutions obtained on changing $L_x =2 $ to $L_x =1$ and the result for different values of $h_w$ and $N$ are shown in Table \ref{tab:c_1d_a_-2_Lx_1}.

\begin{table}[H]
\centering
\caption{Comparison of percentage errors in Performance Index (PI) for different $h_w$ and different number ($N=2 N_s$) of sinusoidal basis functions. Here, we consider the cases for $N_s = N_c$, i.e $N = N_s + N_c = 2 N_s$, with $L_x =1$.}
\label{tab:c_1d_a_-2_Lx_1}
\begin{tabular}{lcl}
\toprule
$h_w$ & $N_s (=N_c)$ & PI error (\%) \\
\midrule
\midrule
0.5  & 8 & $6.3 \times 10^0$ \\ 
0.5 & 16  &  $3.0 \times 10^0$\\ 
0.1  & 8   & $7.7 \times 10^{0}$ \\ 
0.1  & 16 & $5.7 \times 10^{-1}$\\ 
\bottomrule
\end{tabular}%
\end{table}

Further, the Gibbs phenomenon observed in \mfig{fig:c_gibbs_control} can be avoided by using a different basis function other than sinusoidal basis functions. For example, one can use Legendre polynomials. We recall that  Legendre polynomials \( P_n(x) \) are defined as orthonormal with respect to the weight function \( w(x) = 1 \) over the interval \([-1, 1]\), i.e., 
\[
\int_{-1}^{1} P_m(x) P_n(x) \, dx = \begin{cases} 
1 & \text{if } m = n \\
0 & \text{if } m \neq n. 
\end{cases}
\]
The Legendre polynomials are conveniently expressed using the Rodrigues' formula for  \( P_n(x) \) as:
\[
P_n(x) = \frac{1}{2^n n!} \frac{d^n}{dx^n} \left( (x^2 - 1)^n \right).
\]

\begin{table}[H]
\centering
\caption{Comparison of percentage errors in Performance Index (PI) for different $h_w$ and different number ($N$) of Legendre basis functions (with $L_x =1$). }
\label{tab:stability_ham_eig_type_1d_a_}
\begin{tabular}{llll}
\toprule
$h_w$ & $N$ & PI error (\%)  \\
\midrule
\midrule
0.5    & 16  & $6.9 \times 10^{-4}$ \\
0.5    & 20   & $7.4 \times 10^{-5}$  \\
0.5    & 32  & $ 1.7 \times 10^{-3}$  \\
0.05 & 32  & $2.2 \times 10^{-2}$  \\
0.05 & 40   & $2.3 \times 10^{-3}$  \\
\bottomrule
\end{tabular}%
\end{table}

\begin{figure}[H]
\centering
\begin{subfigure}{.5\textwidth}
  \centering
  \includegraphics[width=.85\linewidth]{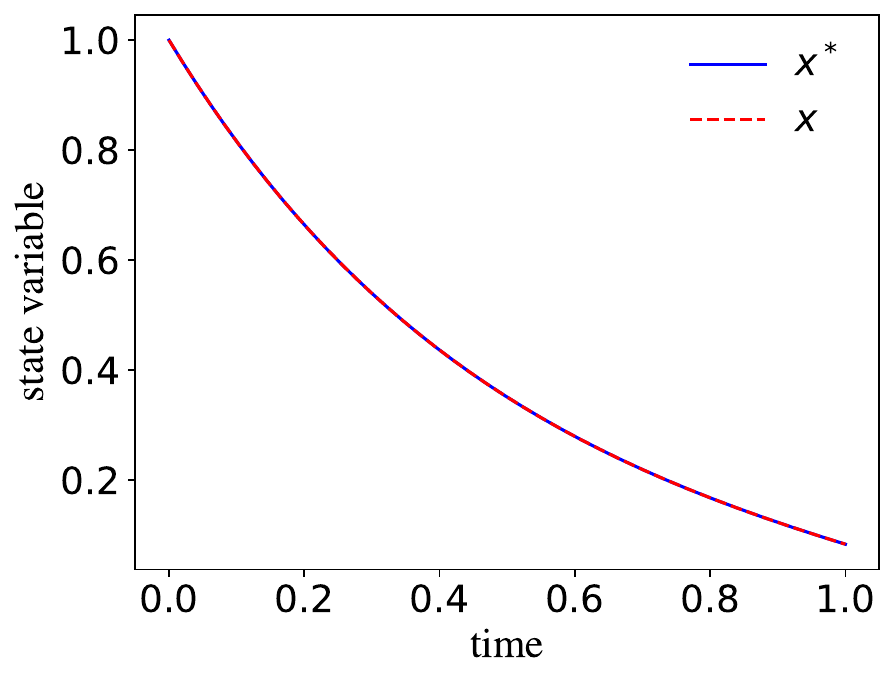}
  \caption{Time evolution of state variable}
  \label{fig:c_a_-2_state_jb}
\end{subfigure}%
\begin{subfigure}{.5\textwidth}
  \centering
  \includegraphics[width=.85\linewidth]{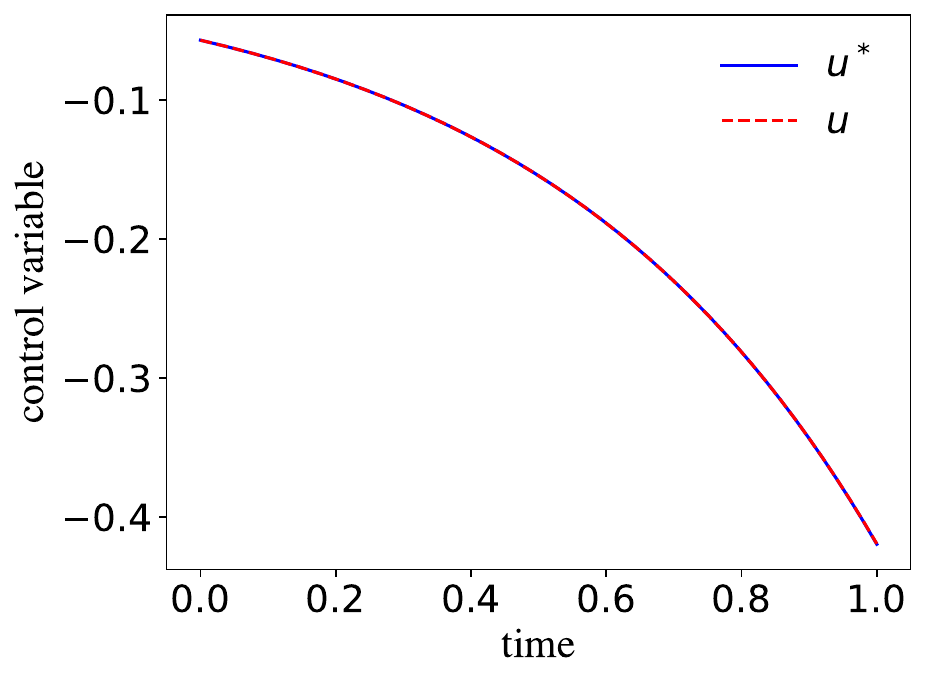}
  \caption{Time evolution of control variable}
  \label{fig:c_a_-2_control_jb}
\end{subfigure}
\caption{Time evolution of state and control variables for Example 3 in \mref{sec:Simulations} (with $N=20$, $h_w = 0.5$ and $L_x = 1$) obtained using Legendre polynomials. Results from the proposed framework (shown in dotted lines) are compared with analytical solutions (shown in solid lines). }
\label{fig:gibbs_jb}
\end{figure}

Indeed, replacing the sinusoidal basis with the Legendre polynomial basis for this problem yields significant improvements in the optimal solutions obtained. Even with a substantially lower number of Legendre basis functions (specifically, \(N=16, \, L_x=1\) and \(h_w = 0.5\)), we achieved a significantly lower percentage PI error of approximately \(6.9 \times 10^{-4} \, \%\), in comparison to the PI error percentage of $6.3  \, \%$ obtained using the sinusoidal basis functions with the same parameters, i.e., \(N=16, \, L_x=1\) and \(h_w = 0.5\)). This also eliminated the Gibbs phenomenon, as seen in \mfig{fig:gibbs_jb} and specifically \mfig{fig:c_a_-2_control_jb}. This reduction in error and elimination of Gibbs phenomenon is notable compared to the results obtained 
using sinusoidal basis functions.
Thus, the choice of basis functions can significantly impact the quality of the obtained results as expected.

\begin{remark} \label{remark:eigen}
\leavevmode
\begin{enumerate}[(a)]
    \item  One expects to obtain the numerical solutions to match with the exact (classical) solutions in the limit $h_w \rightarrow 0$. Besides, one may also expect that for a small but fixed value of $h_w$, increasing the number of basis functions could lead to a better match with the exact (classical) solution. However, we observe that for a fixed $h_w$, increasing the number of basis functions initially improves the approximations, but after a point, increasing the number of basis functions degrades the numerical solution. This was observed for $N=16$ and $N=32$ cases for $h_w=1$ in Table \ref{tab:c_1d_a_-2}, and for $N=20$ and $N=32$ cases for $h_w=1$ in Table \ref{tab:stability_ham_eig_type_1d_a_}.
    \item 
   For a fixed $h_w$, further increasing the value of $N$ may lead to unacceptably high errors in the obtained solutions, and this behavior is related to the nature of eigenvalues of the Hamiltonian matrix. For example, solving the problem in Example 3, using the Legendre basis functions with $h_w = 0.5$  $N = 40$, $L_x =2$ leads to an extremely high error. It is noteworthy that in this case, the eigenvalues of the Hamiltonian matrix with the smallest real part are $-64.46 \, \pm \, 92.07 i$. The presence of highly negative real components of the lowest eigenvalues may lead to an extremely large amplitude for the term $\exp\left(\frac{\lambda_n }{h_w} (t - t_f) \right)$ in the expression for the wave function in \meqref{eq:defpsifullexpansion} (as $t \leq t_f$), leading to failure of the numerical scheme for solving the problem.
   
\end{enumerate}

\end{remark}

\subsection{Example 4}
\label{Sec:example4}
For our next example, we take up the special case of the problem given in \meqref{eq:defL} and \meqref{eq:defxdot}.
We consider the dynamic system described by the differential equation
\begin{equation}
\dot{x} = -x^3 + u, \quad x(0) = \frac{1}{2},
\end{equation}
with the quadratic performance index
\begin{equation}
J = \frac{1}{2} x^2(2) + \frac{1}{2} \int_0^2 \left( x^2 + u^2 \right) dt.
\end{equation}
We seek to determine the control \( u(t) \) that minimizes the cost functional \( J \).
In our notation used in \mref{Sec:Canonical_quantization}, the problem corresponds to the following:
\begin{align}
    \label{eq:1dexfour}
a(x) = - x^3, \quad
b(x) = 1, \quad
c(x)  = 1, \quad
V(x) = \frac{x^2}{2}  \quad
t_0 = 0, \quad
t_f = 2, \quad 
x_0 = \frac{1}{2}.
\end{align}
In this case, we fix the following
\begin{align}
L_x =2,
 \quad
N_s = N_c= 16 \, \, (\text{i.e.},\, N = 32), \quad
h_w = 0.1, \quad R(x, t_f) = 1.
\end{align}
With the above setup, on using our approach described in \mref{Sec:Canonical_quantization}, we obtain a PI of $0.110739$, whereas the PI using MPOPT (an open-source python package to solve multi-stage non-linear optimal control problems  using pseudo-spectral collocation methods \cite{thammisetty2020development})  is $0.11067$. 

\begin{figure}[H]
\centering
\begin{subfigure}{.5\textwidth}
  \centering
  \includegraphics[width=.85\linewidth]{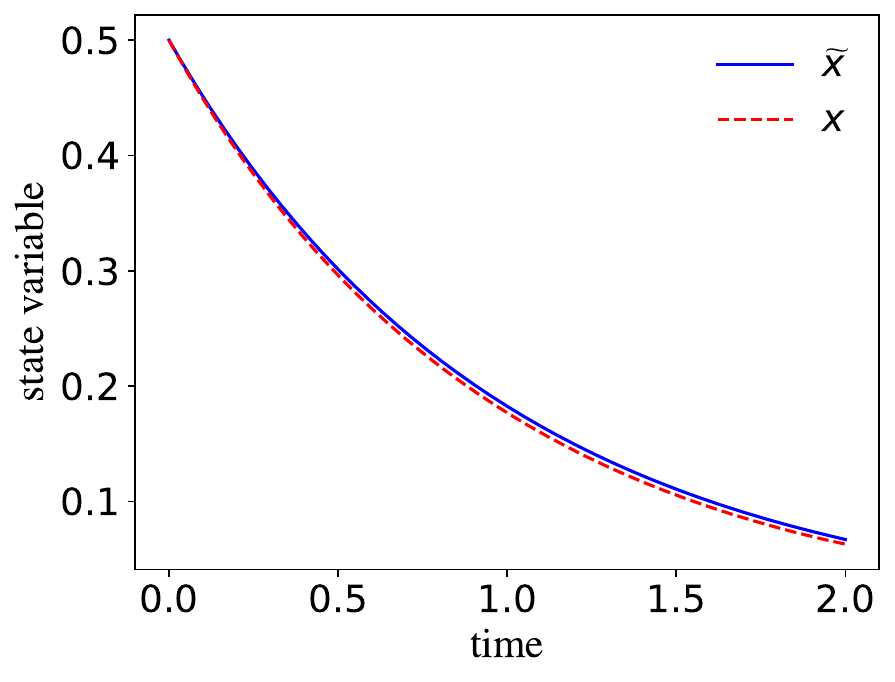}
  \caption{Time evolution of state variable}
  \label{fig:state_a_2_example4}
\end{subfigure}%
\begin{subfigure}{.5\textwidth}
  \centering
  \includegraphics[width=.85\linewidth]{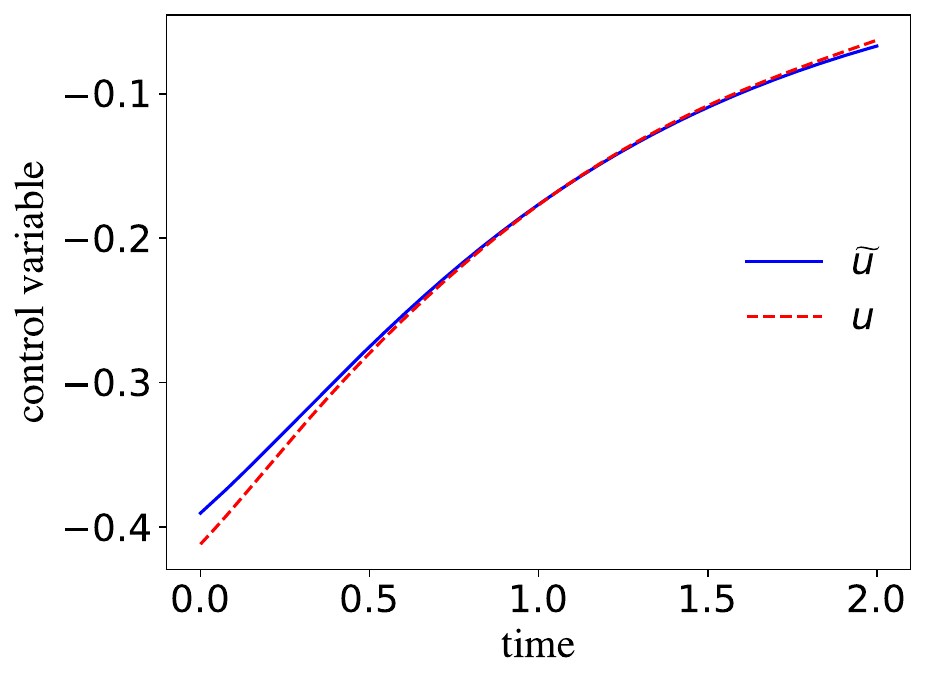}
  \caption{Time evolution of control variable}
  \label{fig:control_a_2_example4}
\end{subfigure}
\caption{Time evolution of state and control variables for Example 4, with $N_s=N_c=16$ (i.e., $N=32$) and $h_w = 0.1$. Results from the proposed framework (dashed lines) are compared with solutions obtained using MPOPT (solid lines).}
\end{figure}

The plots showing the time evolution of the state and control variables, along with the corresponding  solutions obtained using MPOPT, are shown in Figure~\ref{fig:state_a_2_example4} and Figure~\ref{fig:control_a_2_example4} respectively. The plots clearly show that the results from our proposed method closely match the corresponding solutions obtained using MPOPT.

\subsection{Example 5 (2-Dimensional example)}
\label{subsubsec:2d_non_vqa}
Next, we consider a two-dimensional example described below (Ref.~Example 4, \cite{paper4}). Given the system dynamics 
\begin{equation}
    \label{eq:paper4}
    \begin{split}
  \begin{pmatrix}
\dot{x}{_1(t)} \\
\dot{x}{_2(t)} \\
\end{pmatrix} =  \begin{pmatrix}
x_1(t)+u_1(t) \\
u_2(t) \\
\end{pmatrix}, \hspace{0.1cm} t \in (0,1],\\
(x_1(0),x_2(0))=(0.5, -0.5),
    \end{split}
\end{equation}
the  objective is to find control functions \( u_1^*(t) \) and \( u_2^*(t) \) that minimizes the given performance index \( J \), 
\begin{align}
    J = \frac{1}{2}x_1^2(1) + \frac{1}{2} \int_0^1 \lob{u_1(t)^2+ u_2(t)^2+x_2(t)^2} \, dx .
\end{align}
In our formulation (described in \mref{Sec:Canonical_quantization}), we have the following setup, 
\begin{align}
    \label{eq:paper4 reformulation}
b_{ij} = \delta_i^j, \quad m_{ij} =   \delta_i^j \quad a_1 = x_1, \quad a_2 = 0, \quad V(x_1, x_2) = \frac{1}{2}x_2^2, \quad t_f = 1, \quad \Phi( x_1(t_f), x_2(t_f) ) = \frac{1}{2} x_1^2.
\end{align}
Further, in this case, we fix the following
\begin{align}
L_1 = L_2 = 2.5,
 \quad
N_{1,s} = N_{2,s} = N_{1,c} = N_{2,c} = 5, \quad
h_w = 0.25, \quad R(x_1,x_2, t_f) = 1.
\end{align}

\begin{figure}[H]
\centering
\begin{subfigure}{.5\textwidth}
  \centering
  \includegraphics[width=1\linewidth]{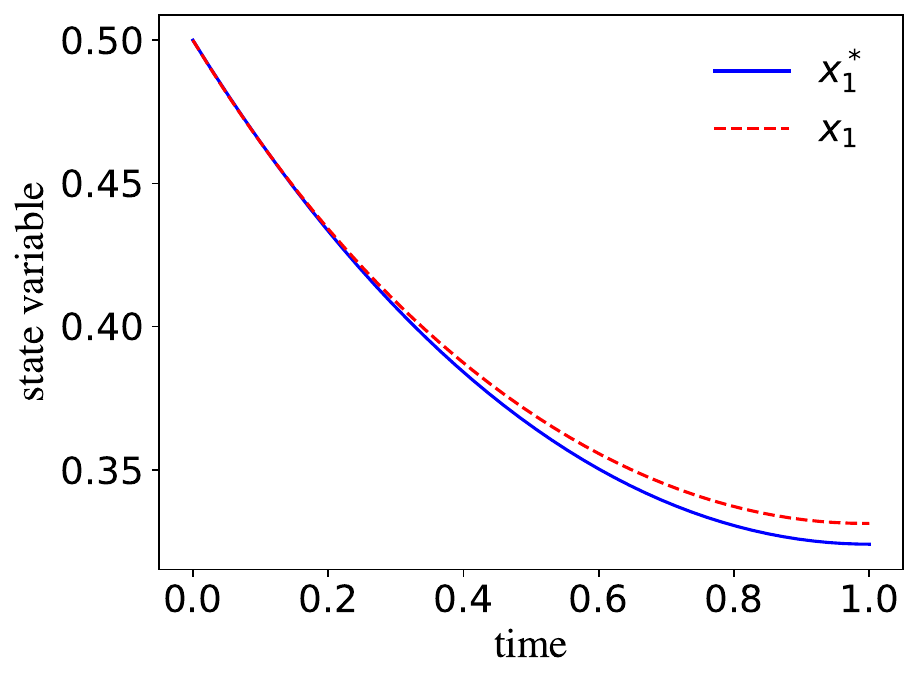}
  \includegraphics[width=1\linewidth]{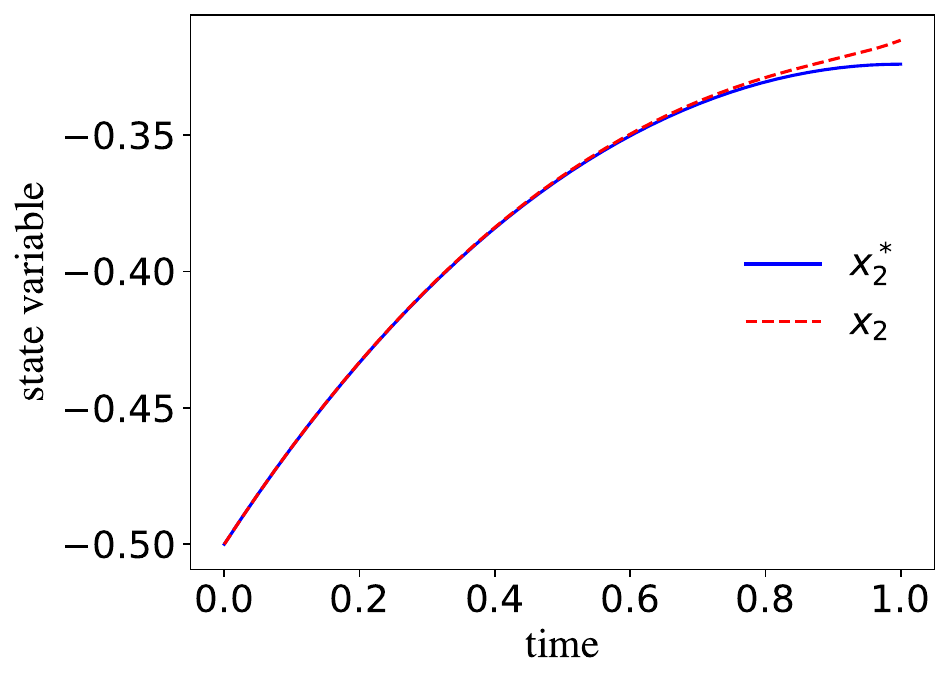}
  \caption{Time evolution of state variables}
  \label{fig:state_paper4}
\end{subfigure}%
\begin{subfigure}{.5\textwidth}
  \centering
  \includegraphics[width=1\linewidth]{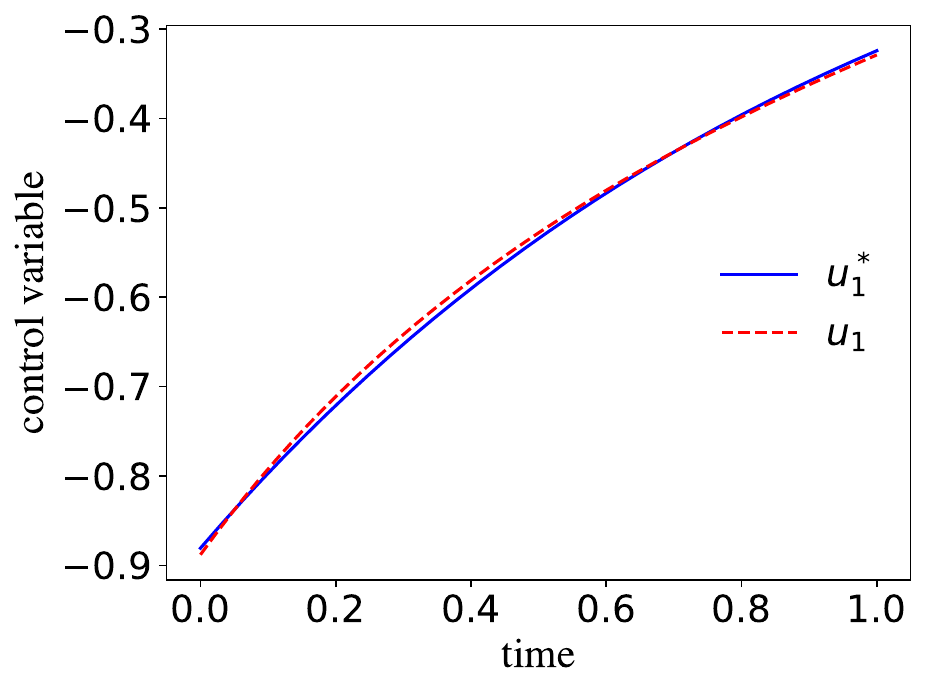}
  \includegraphics[width=1\linewidth]{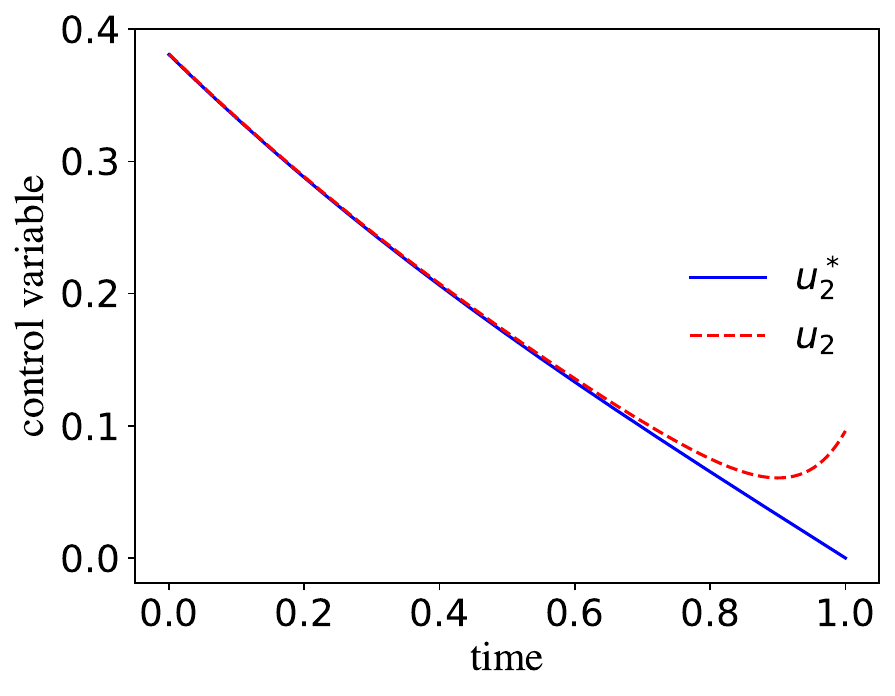}
  \caption{Time evolution of control variables}
  \label{fig:control_paper4}
\end{subfigure}
\caption{Time evolution of state and control variables for the Example 5 in \mref{subsubsec:2d_non_vqa} (with $h_w = 0.25$ and $ N_{1,s} = N_{2,s} = N_{1,c} = N_{2,c} = 5$). Results from the proposed framework (shown in dotted lines) are compared with analytical solutions (shown in solid lines). }
\label{fig:paper4_2d}
\end{figure}

With the above formulation, we obtain the PI of $0.31564$, whereas the correct PI value is $0.3154$. Giving an error of 0.07\%. The graphs of state and control variables are shown in Figure~\ref{fig:state_paper4} and Figure~\ref{fig:control_paper4} respectively.

\subsection{Example 6 (using VQA)}
\label{subsec:vqa_modification}
Next, we consider the problem described in Example 2 in \mref{Sec:example2} and solve it using our framework as before, but now using a slightly modified version of VQA approach for a non-Hermitian system as described in \cite{Xie2024} for computing the eigenvalues of the Hamiltonian matrix. The modifications in the VQA approach include using the ansatz to a form similar to the Efficient-SU2 ansatz proposed by authors in \cite{Kandala_2017}, with fewer entangling layers. This resulted in improved accuracy of solutions obtained for our specific use case.
Further, the ``spectrum scanning'' algorithm in \cite{Xie2024} was modified to include scanning for only the cases where the real part of the eigenvalues are greater than some fixed negative value (it is desirable to have eigenvalues of the Hamiltonian matrix such that they have a positive real part, as as 
noted in Remark \ref{remark:eigen} (b).) 

\begin{figure}[H]
\centering
\begin{subfigure}{.5\textwidth}
  \centering
  \includegraphics[width=.85\linewidth]{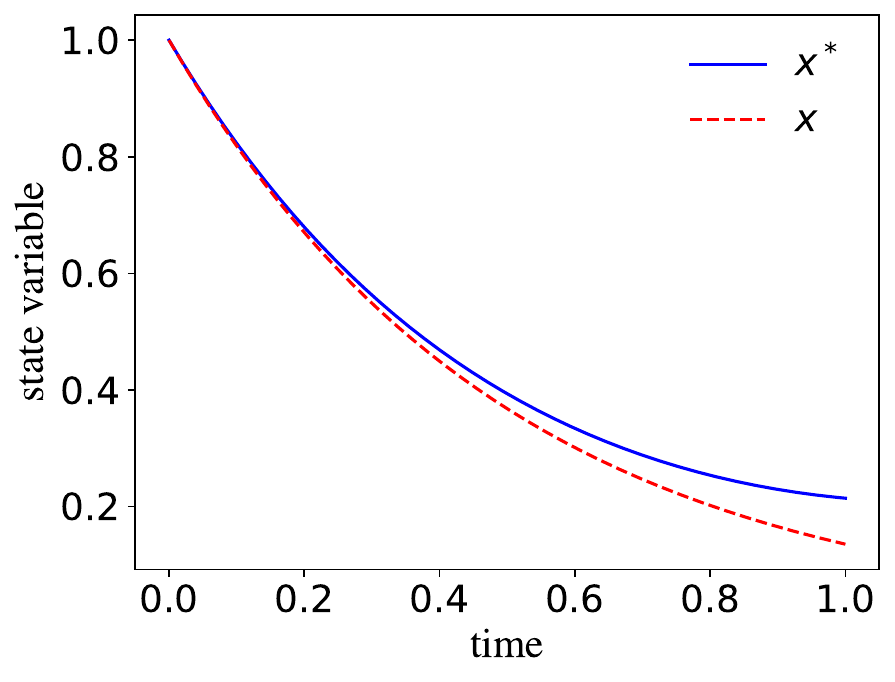}
  \caption{Time evolution of state variable}
  \label{fig:c_vqa_state}
\end{subfigure}%
\begin{subfigure}{.5\textwidth}
  \centering
  \includegraphics[width=.85\linewidth]{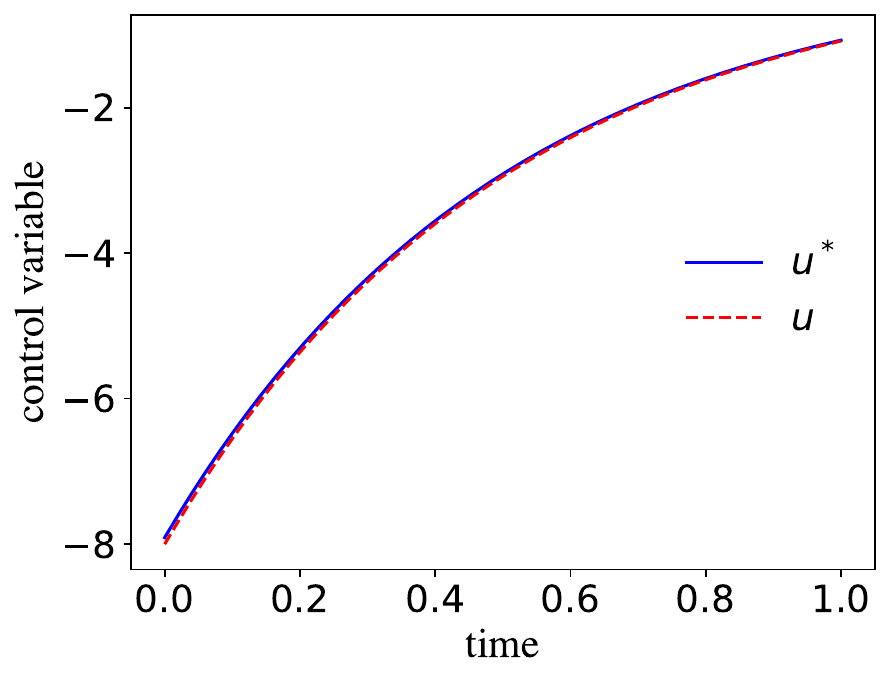}
  \caption{Time evolution of control variable}
  \label{fig:c_vqa_control}
\end{subfigure}
\caption{Time evolution of state and control variables for Example 2 in \mref{sec:Simulations} (with $N_1=N_2=8$ and $h_w = 2$). Results from the proposed framework based on VQA (shown in dotted lines) are compared with analytical solutions (shown in solid lines). }
\label{fig:c_vqa}
\end{figure}

On using our framework described in \mref{Sec:Canonical_quantization} based on the VQA method as described in the previous paragraph, we obtained the PI of $7.9451$ compared to the actual PI of $7.91304$, resulting in an error of $0.4 \, \%$. 
The time evolution of state and control variables, along with the respective correct analytic solutions, are shown in  Figure~\ref{fig:c_vqa_state} and Figure~\ref{fig:c_vqa_control}, respectively, and they are in good agreement. The results could be improved further by employing improved variational quantum eigensolvers for non-Hermitian systems.

\section{Conclusion}\label{sec:conclusion}

In this paper, we proposed a novel variational quantum approach for solving a class of nonlinear optimal control problems. This method integrates Dirac's approach for the canonical quantization of dynamical systems with the solution of the ground state of the resulting non-Hermitian Hamiltonian via a variational quantum eigensolver (VQE).

We introduced a new perspective on the Dirac bracket formulation for generalized Hamiltonian dynamics in the presence of constraints, providing clear motivation and illustrative examples. Additionally, we explored the structural properties of Dirac brackets within the context of multidimensional constrained optimization problems.

Our proposed approach for solving a class of nonlinear optimal control problems uses the VQE to determine the (left or right) eigenstate and corresponding eigenvalue (associated with ground state energy) of a non-Hermitian Hamiltonian, employing a slightly modified version of the approach proposed in \cite{Xie2024}. Assuming access to an ideal VQE, our formulation produced excellent results, as demonstrated by our selected computational examples. Furthermore, our method also performed well when combined with the VQE approach for non-Hermitian Hamiltonian systems described in \cite{Xie2024}. These results could be further enhanced by employing more advanced variational quantum eigensolvers for non-Hermitian systems.

Our formulation with VQE effectively addresses the challenges associated with a wide range of optimal control problems, particularly in high-dimensional scenarios. When compared to standard classical approaches for solving optimal control problems, our quantum-based method shows significant promise and offers an alternative for tackling complex, high-dimensional optimization challenges.

\section*{Acknowledgement}
The authors would like to thank all the authors of \cite{Xie2024} and especially  Dr.~Xu-Dan Xie for sharing the code relevant to VQA for non-Hermitian systems.


\end{document}